\documentclass[a4paper, 9pt]{article}

\usepackage{amssymb}
\usepackage[table, dvipsnames]{xcolor}
\usepackage{booktabs}
\usepackage{chemformula}                        
\usepackage{siunitx}                            
\usepackage[colorlinks=true, linkcolor=black, citecolor=ForestGreen, urlcolor=cyan]{hyperref}
\usepackage{float}                              
\usepackage[capitalise, noabbrev]{cleveref}     
\usepackage{authblk}                            
\usepackage[labelfont=bf,font=small]{caption}   
\usepackage{csvsimple}                          
\usepackage[UKenglish]{babel}                   
\usepackage{graphicx}
\usepackage{doi}
\usepackage{changepage}

\usepackage[super,comma,sort&compress]{natbib}

\usepackage{microtype}
\usepackage[notextcomp]{stix2}
\usepackage[left=3.5cm, right=3.5cm, top=2.5cm, bottom=2.5cm]{geometry}
\usepackage{parskip}

\usepackage{titlesec}
\titleformat{\subsection}{\normalfont\large\bfseries}{\thesubsection}{1em}{}

\crefname{figure}{Fig.}{Figs.}
\Crefname{figure}{Fig.}{Figs.}

\newfloat{supfigure}{htbp}{lop}[section]
\floatname{supfigure}{Supplementary Figure}

\crefname{supfigure}{Supplementary Fig.}{Supplementary Figs.}
\Crefname{supfigure}{Supplementary Fig.}{Supplementary Figs.}

\newfloat{suptable}{htbp}{lop}[section]
\floatname{suptable}{Supplementary Table}

\crefname{suptable}{Supplementary Tab.}{Supplementary Tabs.}
\Crefname{suptable}{Supplementary Tab.}{Supplementary Tabs.}

\DeclareMathOperator*{\argmax}{arg\,max}

\title{Little to lose: the case for a robust European green hydrogen strategy}

\author[1]{Koen van Greevenbroek\thanks{Corresponding author, \url{koen.v.greevenbroek@uit.no}}}
\author[2]{Johannes Schmidt}
\author[3]{Marianne Zeyringer}
\author[1]{Alexander Horsch}

\affil[1]{Department of Computer Science, UiT The Arctic University of Norway, Postboks 6050 Langnes, 9037 Tromsø, Norway}
\affil[2]{Institute of Sustainable Economic Development, University of Natural Resources and Life Sciences, Feistmantelstraße 4, 1180 Vienna, Austria}
\affil[3]{Department of Technology Systems, University of Oslo, P.O. Box 70, 2027 Kjeller, Norway}

\date{\today}

\begin{document}

\maketitle

\begin{center}
\begin{minipage}{0.7\textwidth}
\textbf{Summary} \\
The EU targets 10 Mt of green hydrogen production by 2030, but has not committed to targets for 2040.
Green hydrogen competes with carbon capture and storage, biomass, imports as well as direct electrification in reaching emissions reductions; earlier studies have demonstrated the great uncertainty in future cost-optimal development of green hydrogen.
In spite of this, we show that Europe risks little by setting green hydrogen production targets at around 25 Mt by 2040.
Employing an extensive scenario analysis combined with novel near-optimal techniques, we find that this target results in systems that are within 10\% of cost-optimal in all considered scenarios with current day biomass availability and baseline transportation electrification.
Setting concrete targets is important in order to resolve significant uncertainty which hampers investments.
Targeting green hydrogen reduces the dependence on carbon capture and storage and green fuel imports, making for a more robust European climate strategy.
\end{minipage}
\end{center}

\vspace{3ex}

As Europe undergoes the transition to net-zero emissions by 2050, hydrogen is seen as a key technology enabling a shift away from fossil fuels.
The European Union has set a target of \qty{10}{Mt/a} green hydrogen production and imports each by 2030 \cite{europeancommission-2022}.
Beyond 2030, however, targets for scaling up green hydrogen are missing even as the EU has proposed a 90\% emissions reduction by 2040 \cite{europeancommission-2024b} and committed to net zero emissions by 2050 by law \cite{eu-climate-law-2021}.
This is characteristic of the fundamental uncertainty around the future role of green hydrogen production in Europe.
In this study, we present the case for committing to European green hydrogen targets despite this uncertainty, while keeping the risk of cost overruns and missing climate targets low.

In net-zero energy systems, green hydrogen (which we take to be any hydrogen produced by electrolysis in this study) is expected to serve primarily as a feedstock for synthetic fuel production, as a fuel in the transportation sector and as a feedstock and heat source in industry \cite{beres-nijs-ea-2024,seck-hache-ea-2022,neumann-zeyen-ea-2023}.
In these roles, provided direct electrification is not possible, green hydrogen competes with fossil fuels combined with carbon capture and storage (CCS) \cite{mignone-clarke-ea-2024}, green fuel imports \cite{neumann-hampp-ea-2024} as well as biomass \cite{ganter-gabrielli-ea-2024,blanco-nijs-ea-2018}.
While estimates of biomass availability by 2050 vary \cite{wu-muller-ea-2022,millinger-hedenus-ea-2023}, most studies \cite{pickering-lombardi-ea-2022,neumann-zeyen-ea-2023,zeyen-victoria-ea-2023,beres-nijs-ea-2024,fleiter-fragoso-ea-2024} in a European context use conservative values of around 1200 TWh solid biomass available per year \cite{ruiz-nijs-ea-2019}, which is close to current values \cite{europeancommission-directorate-generalforenergy-2024}.

Development of CCS and green fuel imports, however, is subject to much greater uncertainty.
Green fuel imports (here: carbon-neutral hydrogen, ammonia, methanol, synthetic gas and oil imports) have the potential for reducing total system costs \cite{neumann-hampp-ea-2024,galimova-ram-ea-2023}, but global supply is currently limited, and is expected to stay scarce for the foreseeable future \cite{muller-riemer-ea-2024,odenweller-ueckerdt-ea-2022,odenweller-ueckerdt-2025}.
Moreover, a continued reliance on energy imports could negate the resilience to short-term supply shocks that European decarbonisation promises \cite{europeancommission-2024}, though green hydrogen may also entail complicated, global supply chains \cite{jacopomariapepe-dawudansari-ea-2023}.

\begin{figure}
  \centering
  \begin{adjustwidth}{-1.5cm}{-1.5cm}
    \includegraphics{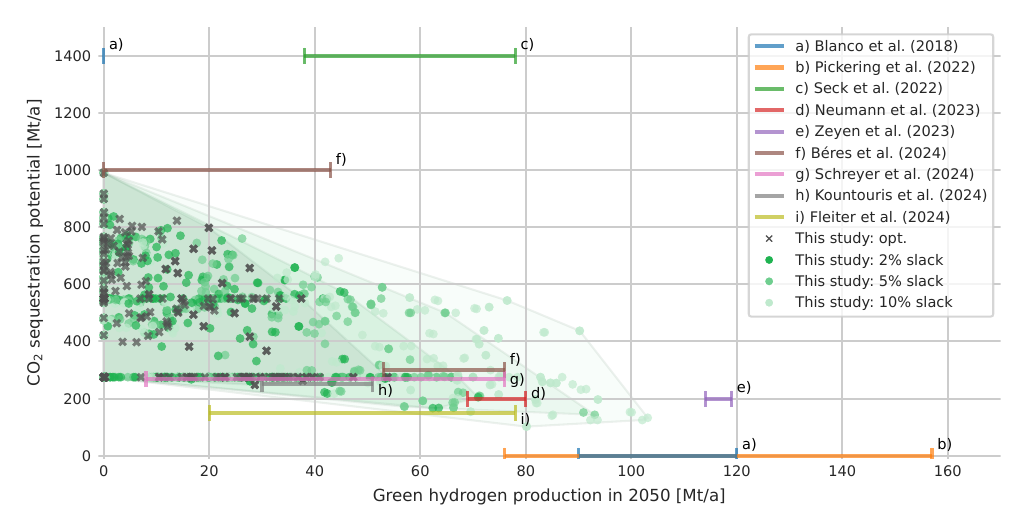}
    \caption[Ranges of green hydrogen production in 2050 found in this study and previous studies]{
      Ranges of green hydrogen production in 2050 found in this and previous studies \cite{blanco-nijs-ea-2018,pickering-lombardi-ea-2022,seck-hache-ea-2022,neumann-zeyen-ea-2023,zeyen-victoria-ea-2023,beres-nijs-ea-2024,schreyer-ueckerdt-ea-2024,kountouris-bramstoft-ea-2024,fleiter-fragoso-ea-2024}, plotted against the amount of \ch{CO2} sequestration in each study. 
      Blanco et al.\ and Béres et al.\ contain different scenarios with different \ch{CO2} sequestration potentials which are plotted here with separate line segments.
      For each study, the plotted \ch{CO2} sequestration level represents either the actual amount of sequestration found in the study (for studies a--e \& h) or the upper limit on sequestration (for studies f, g \& i) in case the actual amount of sequestration is not given.
      In studies c, d \& e, actual \ch{CO2} sequestration coincides with the upper limit imposed.
      For each given level of \ch{CO2} sequestration, we plot only the range from the least to the greatest amount of green hydrogen production found in the study results; the number of different scenarios/results varies from a few to hundreds between the different studies.

      Results from the present study are plotted with green dots, shaded by (near-) optimality.
      For near-optimal results, both max- and minimisations of green hydrogen are included.
      With a total of 216 scenarios and three slack levels (with a minimisation and maximisation for each slack level), this makes for $216 \cdot (1 + 2 \cdot 3) = 1512$ data points, though some points are missing due to failed optimisations.
      For easier interpretation, the convex hulls of the four sets of points with different slack levels (optimal, 2\%, 5\%, 10\%) are also drawn.
      Only results for 2050 are included.

      All previous studies work with a net-zero emissions constraint given the 2050 time horizon, except \citeauthor{blanco-nijs-ea-2018} who assume a 95\% reduction in emissions instead.
      We limit the inclusion of studies to only those modelling the whole European energy system, including electricity, heating, transportation and industry sectors; whether (net) emissions from agriculture and land use, land use change and forestry (LULUCF) are included, however, varies between studies.
      Moreover, while all studies cover at least the EU, some also cover additional European countries (like the present study).
      See also \cref{suptab:green-h2-lit-review} for a table with the same data on previous studies, including additional brief comments on the methodologies used.
      While we report quantities of hydrogen in millions of tonnes (Mt), this can be converted to energy content using the lower heating value of \qty{33.3}{kW/kg\ch{H2}}, such that $\qty{1}{Mt} = \qty{33.3}{TWh}$ for hydrogen and $\qty{1000}{TWh} \approx \qty{30}{Mt}$.
    }
    \label{fig:lit-review-ranges}
  \end{adjustwidth}
\end{figure}

The prospects for fossil fuels and CCS towards 2050 are contentious.
While the EU has committed to a target of \qty{50}{Mt/a} of \ch{CO2} sequestration (permanent geological storage) by 2030 through the Net Zero Industry Act \cite{eu-net-zero-industry-2024}, there is wide disagreement among energy systems studies on how this number might scale up in the coming decades (\cref{fig:lit-review-ranges}).
The impact assessment for the EU 2040 climate target \cite{europeancommission-2024} suggests the need for some \qty{250}{Mt/a} sequestration capacity by 2040.
Apart from technological challenges in scaling up CCS \cite{kazlou-cherp-ea-2024}, its large-scale use also relies on the continued, unsustainable use of depleting fossil fuel reserves as well as a finite \ch{CO2} storage capacity \cite{anthonsen-christensen-2021}.
These factors, among others \cite{quarton-tlili-ea-2020}, conspire to produce wide ranges of projected (optimal) green hydrogen production levels in Europe by 2050 (\cref{suptab:green-h2-lit-review}).

Net-zero energy systems entirely without green hydrogen production have been shown to be feasible and even cost-optimal in some previous studies \cite{beres-nijs-ea-2024,blanco-nijs-ea-2018}, but only with 1000--\qty{1400}{Mt} of \ch{CO2} sequestration per year (\cref{fig:lit-review-ranges}).
Especially synthetic fuel production from green hydrogen is out-competed by fossil fuels under liberal assumptions on \ch{CO2} sequestration potential \cite{victoria-zeyen-ea-2022}; the sequestered \ch{CO2} is primarily obtained from point source carbon capture (from biomass or fossil fuel combustion) and possibly direct air capture to compensate for diffuse sources of fossil emissions, such as in aviation.
The European Environmental Agency, meanwhile, uses a \qty{500}{Mt/a} feasibility threshold for \ch{CO2} sequestration in the 2040 climate target impact assessment \cite{europeanenvironmentagency-2023a}.

Despite doubts around the profitability of green hydrogen towards 2050, it may nonetheless be advantageous to set concrete targets for scaling up its production.
If green hydrogen does not take off, achieving net-zero emissions in hard-to-electrify sectors will require a large expansion of CCS capacity and/or green fuel imports.
Thus, scaling up green hydrogen production can make the transition to net-zero emissions more robust against technological uncertainty and external shocks.
Moreover, clearly articulated visions for a future energy system are to be preferred in transitions \cite{sgouridis-kimmich-ea-2022a}, reducing future uncertainty and recognising path-dependencies and learning-by-doing \cite{zeyen-victoria-ea-2023}.

We show that wide ranges of green hydrogen production are possible when relaxing cost-optimality slightly.
Across a systematic, large-scale scenario exploration, we find that allowing a 5\% increase in total energy system costs opens up a difference of \qty{38}{Mt/a} between minimum and maximum feasible green hydrogen production in 2050 on average.
Crucially, a moderate target such as a total yearly production of \qty{25}{Mt} by 2040 is found to be feasible and near-optimal (no more than 10\% more expensive than cost-optimal) across all scenarios with baseline assumption on biomass availability and transportation electrification.
Thus, not only would such a target secure the energy transition against a variety of uncertainties and shocks: it is highly unlikely to lead to severe cost overruns.
With the total cost overrun bounded by 10\% of the system cost, amounting to some \qty{91}{bn~EUR} on average, we find reaching the target would require only \qty{13.7}{bn~EUR/a} of green hydrogen subsidies on average.
Europe has little to lose by committing to a moderate green hydrogen production target.

Green hydrogen development is tied directly to the future of fossil fuels as well as renewable electricity, representing either a competitor or an opportunity for different interest groups.
Cost-optimal modelling results with a ``central planning'' approach may not capture system designs that are politically more viable but slightly more costly \cite{trutnevyte-2016}, motivating our focus on near-optimal methods, also known as \emph{Modelling to Generate Alternatives} (MGA) \cite{decarolis-2011}.
With such methods, it is also possible to generate options that are feasible and near-optimal across a number of scenarios \cite{vangreevenbroek-grochowicz-ea-2023}.

Up to this moment, however, near-optimal methods have only been developed for single-horizon optimisations \cite{millinger-hedenus-ea-2023,pickering-lombardi-ea-2022,decarolis-2011,neumann-brown-2021,neumann-brown-2023,grochowicz-vangreevenbroek-ea-2023} (including overlapping independent results from different planning horizons \cite{esser-finke-ea-2024}) or multi-horizon optimisations with perfect foresight over the planning horizon \cite{price-keppo-2017,sinha-venkatesh-ea-2024}.
We therefore extend the state of the art by developing for the first time near-optimal methods in the context of multi-horizon optimisations with myopic planning foresight.
These methods are well suited to ``min-max'' applications of MGA, though may not generalise fully to other types of near-optimal exploration.
In our context, the method allows us to study the temporal dynamics of the energy transition while limiting computational complexity and running the underlying energy system model at the high temporal and spatial resolution necessary to capture renewable variability and long-term storage; optimisations with perfect foresight across planning horizons are computationally intractable at this scale.
As opposed to single-horizon optimisation, the multi-horizon approach allows for the study of lock-in effects and stranded assets.

Our application of new multi-horizon near-optimal techniques allows us for the first time to investigate robust pathways of European green hydrogen production; a first step on the way to a robust production target for 2040 and beyond.
Neumann et.~al.~\cite{neumann-brown-2023} and Pickering et.~al.~\cite{pickering-lombardi-ea-2022} employ near-optimal methods to generate large numbers of solutions, but do not consider green hydrogen production explicitly as a variable of interest, only consider a 2050 planning horizon, and only consider \ch{CO2} sequestration potentials of \qty{200}{Mt/a} and \qty{0}{Mt/a}, respectively.
Other research on European hydrogen production towards 2050 considers only cost-optimal pathways, thus underestimating the true ranges of viable production levels.
Our study also represents the most systematic and extensive scenario analysis on the question of European green hydrogen to date.
We obtain results over combinations of different assumptions across categories representing CCS potential, biomass availability, presence of imports, electrolysis costs, transportation electrification and weather year.
Resulting in a total of 216 scenarios, we can show exactly which factors are the most significant for the future of green hydrogen in Europe.

\subsection*{Mapping out green hydrogen production pathways}

\begin{figure}
  \centering
  \begin{adjustwidth}{-1.5cm}{-1.5cm}
    \includegraphics[width=17cm]{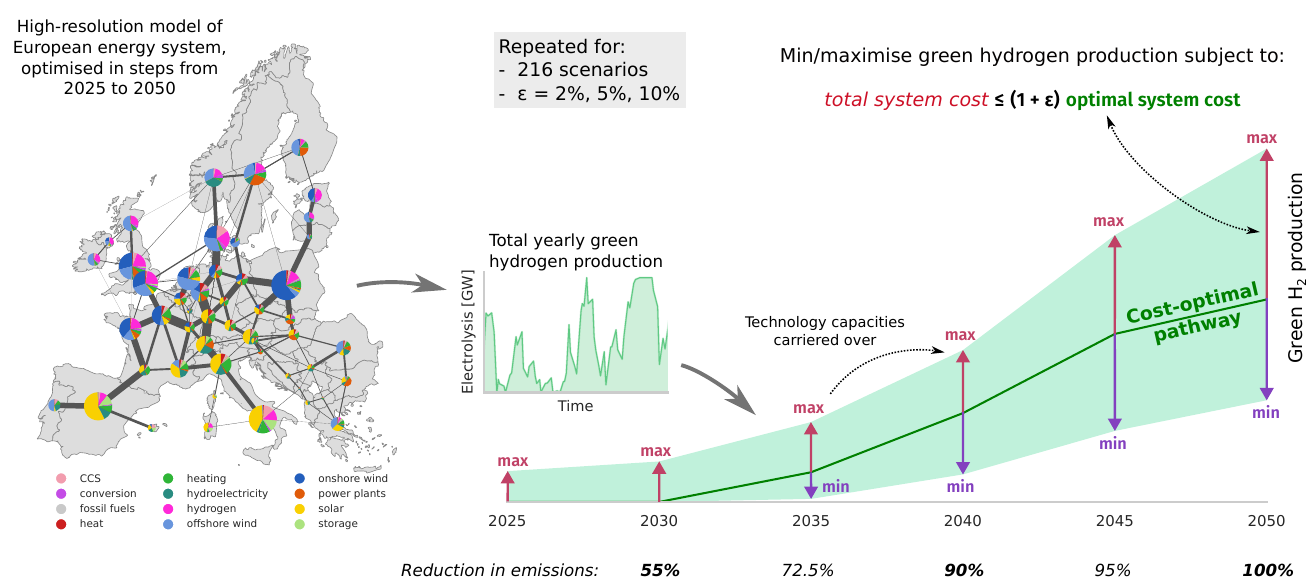}
    \caption{
      Illustration of model layout and the sequences of optimisations used to obtain ranges of green hydrogen production.
      On the left, an illustration of the spatial extent and resolution of the model used in this study.
      Note that the modelling region includes the EU (minus Cyprus and Malta) in addition to the UK, Switzerland, Norway and most Balkan countries.
      For an exact listing of which technologies are included in which legend categories, see the plotting code (Code Availability).
      In the middle, the cost-optimal pathway, consisting of a sequence of cost-optimisations at 5-year intervals, where capacities from each optimisation are carried over to the next (minus capacities that are phased out).
      Blue and red arrows illustrate the minimisations and maximisations (respectively) of green hydrogen at each time horizon; total system costs are not allowed to exceed $(1+\varepsilon)$ times the optimal system cost for each time horizon.
      This sequence of optimisations is repeated for each of the 216 considered scenarios, and with $\varepsilon=2\%,5\%,10\%$.
    }
    \label{fig:methodology}
  \end{adjustwidth}
\end{figure}

The subsequent results are based on a large number of energy system model optimisations covering the period of 2025--2050 in 5-year steps; our modelling region includes most of the EU in addition to the UK, Switzerland, Norway and most Balkan countries.
A model-wide cap on \ch{CO2} emissions is gradually tightened in line with EU targets: 55\%, 90\% and 100\% reductions by 2030, 2040 and 2050, respectively.
In order to fully capture the potential roles played by hydrogen, the model (based on PyPSA-Eur \cite{PyPSAEur,PyPSAEurSec}) includes representations of the electricity, gas, heating, transportation and industry sectors, and is run at high spatial and temporal resolution (Methods).
Green hydrogen production is minimised and maximised subject to total system cost increase limits of $2\%$, $5\%$ and $10\%$ at each time horizon and for each scenario separately (\cref{fig:methodology}).

We repeat the optimisations for a total of 216 scenarios, representing all possible combinations of different levels in six settings:
\begin{enumerate}
\item CCS potential: (a), (b), (c).
\item Biomass potential: (a), (b), (c).
\item Green fuel import potential: (a), (b).
\item Electrolyser capital cost: (a), (b).
\item Transportation electrification: (a), (b), (c).
\item Weather year: (a), (b).
\end{enumerate}

The details of the scenario assumptions can be found in the Methods section; for each setting our assumptions range from pessimistic (a) to optimistic (b or c).
The CCS potential setting consists of a combination of sequestration potential, marginal cost of sequestration and capital cost of carbon capture infrastructure.
Biomass potentials are derived from the ENSPRESO database \cite{ruiz-nijs-ea-2019}.
Green fuel imports are either limited in volume to less than domestic green hydrogen production, or unlimited.
Optimistic and pessimistic transportation settings shift electrification rates forwards and backwards by 5 years, respectively.

\subsection*{Large differences in the importance of green hydrogen for 2040 climate target}

\begin{figure}
  \centering
  \begin{adjustwidth}{-1.5cm}{-1.5cm}
    \includegraphics{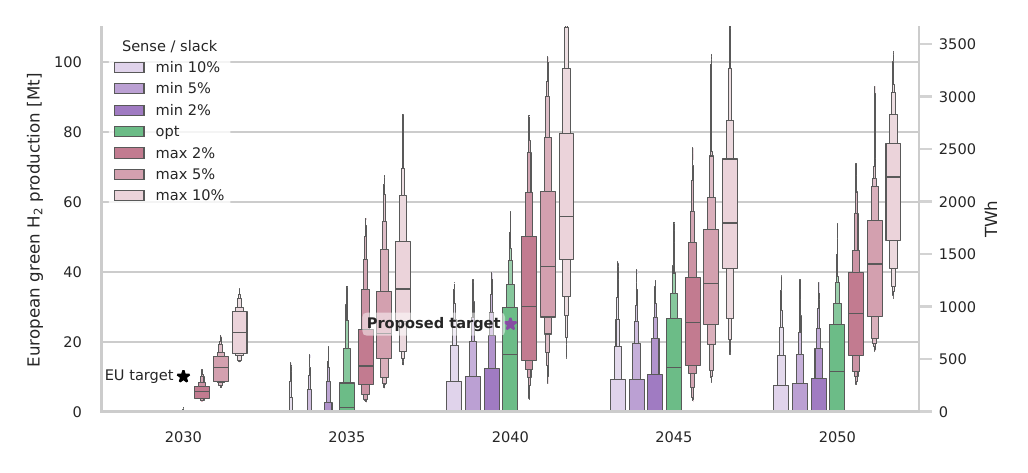}
    \caption{
      The range of pathways for European green hydrogen production across the scenarios studied.
      Shown are the cost-optimal ranges (green) as well as ranges under green hydrogen minimisation (blue) and maximisation (red).
      For the min- and maximisations, the ranges are shown for 2\%, 5\% and 10\% total system cost slack levels.
      For each category, we use a letter-value plot \cite{letter-value-plot} to visualise the distribution of green hydrogen production levels in the 216 different scenarios.
      The lines in the middle indicate the median, and successive boxes mark the 0.25--0.75 quantiles (containing 50\% of the data points), the 0.125--0.875 quantiles (containing 25\% of the data), etc.
      While we give hydrogen production figures in Mt/a and TWh/a, \cref{supfig:h2-vs-elec-cap} shows the relationship with installed electrolyser capacity, which reaches about 750--\qty{1000}{GW} or more for \qty{80}{Mt} of annual green hydrogen production.
      The peak in hydrogen production in 2040 is primarily caused by remaining internal combustion engine cars in combination with the 90\% emissions reduction target causing significant demand for synthetic liquid fuels, which subsequently drops with further land transport electrification.
    }
    \label{fig:pathways}
  \end{adjustwidth}
\end{figure}

Different scenarios lead to a large variety of cost-optimal green hydrogen production pathways, with an interquantile range of 0--\qty{30}{Mt/a} already by 2040 (\cref{fig:pathways}).
Sensitivities to individual scenario settings are explored in \cref{tab:sensitivity}.
Allowing for just a 2\% total system cost increase, the range widens significantly, with different scenarios leading to maximum levels of green hydrogen production with an interquantile range of 15--\qty{50}{Mt/a}.
At a 10\% total system cost slack, the upper end of the range extends to \qty{100}{Mt/a} in some scenarios.

Green hydrogen production is nearly absent in all cost-optimal scenarios for 2030 --- it is not necessary in order to reach a 55\% reduction in emissions.
While this means that the EU target to produce \qty{10}{Mt/a} by that year will require financial incentives in any case, said target could still kick-start growth and technological learning in the sector \cite{zeyen-victoria-ea-2023}.
Grey hydrogen still supplied most demand in 2030, but is largely phased out by 2040; in some scenarios, minimising green hydrogen production leads to the use of solid biomass for hydrogen production in 2035 and 2040 (\cref{supfig:h2prod-by-tech}).
Blue hydrogen does not play a large role in most scenarios, but experiences an uptake after 2040 in some scenarios (\cref{supfig:h2prod-by-tech}).
On average across all cost-optimal scenarios, 79.7\% of all European production is green (i.e. electrolytic) by 2040, falling slightly to 60.9\% by 2050.

In most scenarios, green hydrogen production peaks already in 2040, largely driven by a strong demand for synthetic fuels in the land transportation sector.
This synthetic fuel demand is in turn induced by the EU's target of a 90\% emissions reduction by 2040, while a significant number of internal combustion engine cars are still expected to be operational in 2040 (Methods).
The land transportation sector is still expected to dominate demand for oil in 2040, only ceding to aviation by 2050 (\cref{supfig:demand-fixed-major}).
Supporting synthetic fuel production, we see massive build-out of solar and wind power, with median installed solar capacity of approximately \qty{1430}{GW} and wind (on- and offshore) of \qty{1110}{GW} by 2040 (\cref{supfig:renewables}).
Imports of green fuels also peak already in 2040 (\cref{supfig:imports}).
Cost-optimal pathways install about \qty{140}{GW} (at the median) of electrolysers already in 2040 (\cref{supfig:elec-cap}), with some maximisation-pathways installing more than \qty{1000}{GW}.
Synthetic fuel needs a source of carbon as well as hydrogen; use of biomass with carbon capture is the preferred pathway in most scenarios.
Still, a quarter of cost-optimal pathways see direct air capture of more than \qty{50}{Mt\ch{CO2}} by 2040; this number rises to more than \qty{100}{Mt\ch{CO2}} for pathways maximising green hydrogen production (\cref{supfig:DAC}).
Available geological storage of \ch{CO2}, limited in 2040 to 125, 250 or \qty{500}{Mt/a} depending on the scenario, is fully used in almost every case (\cref{supfig:sequestration}), but is generally negatively correlated with green hydrogen production (\cref{supfig:h2-vs-co2seq}).

Using green synthetic fuels in internal combustion engines is a much less efficient way of decarbonising the land transport sector than direct electrification --- a peak and subsequent decline in green hydrogen production could also lead to stranded electrolyser assets.
Indeed, if land transportation electrification is sped up by 5 years compared to the baseline transition pathway, the green hydrogen production peak in 2040 is almost entirely subdued, with the median cost-optimal production level at only \qty{6}{Mt} (\cref{supfig:h2prod-by-transportation}).
These results indicate that a 90\% emissions reduction by 2040 would be much more easily achieved if direct electrification is sped up.

\subsection*{Uncertainty around green hydrogen extends to fossil fuels and beyond}

\begin{figure}
  \centering
  \begin{adjustwidth}{-1.5cm}{-1.5cm}
    \includegraphics{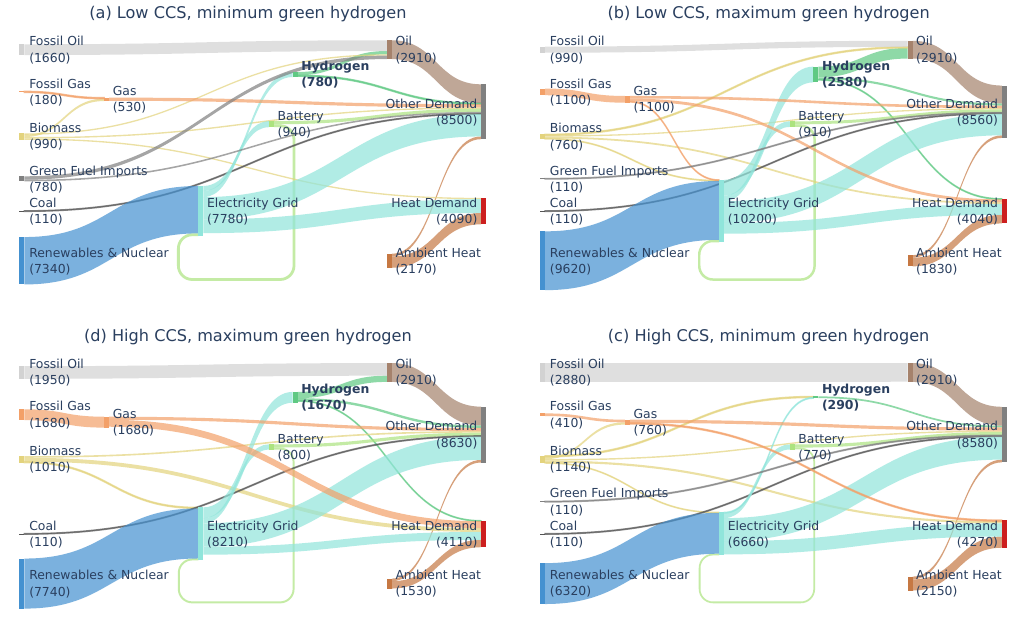}
    \caption{
      Comparison of energy flows in two different scenarios (top and bottom) under minimisation and maximisation of green hydrogen production (left and right, respectively).
      Total energy input to each node is displayed in TWh, rounded to the nearest multiple of 10.
      Connections with a value of less than \qty{50}{TWh} are not displayed.
      This figure displays yearly total energy flows at the 2040 planning horizon, with a 5\% total system cost slack for green hydrogen min/maximisation.
      The ``Low CCS'' and ``High CCS'' labels correspond to levels (a) and (c) of the CCS setting, respectively (Methods). 
      Apart from differing in CCS assumptions, the systems are all comparable in having restricted green fuel imports, optimistic electrolyser cost assumptions ($-50\%$ capital cost), medium biomass availability, a baseline land transport electrification pathway and are modelled over the 1987 weather year.
      Green hydrogen production amounts to \qty{23.4}{Mt} (a), \qty{77.5}{Mt} (b), \qty{3.4}{Mt} (c) and \qty{50.0}{Mt} (d) in the respective panels (recalling that $\qty{1}{Mt} = \qty{33.3}{TWh}$ using the lower heating value).
      Note that total demand is slightly different in the different panels since part of the demand is endogenously modelled and can change between optimisations, such as direct air capture technology.
      The ``Heat Demand'' category includes all heating demand from the residential and services sector; industrial process heat and all other energy demand is gathered under ``Other Demand''.
      Separating out residential and services heat demand makes the differences in the role of natural gas for heating between the different panels visible.
      For a detailed exposition of energy demand, see the Methods, as well as \cref{supfig:demand-fixed-major,supfig:demand-fixed-minor} for an overview over exogenously fixed energy demand.
      Energy losses are not shown explicitly here for the sake of simplicity; they are in the range of 890--\qty{2690}{TWh}, with especially losses in the use of electricity (in large part due to green hydrogen production) varying between 200 and \qty{1370}{TWh}.
      Energy flows for the same scenarios in 2050 are shown in \cref{supfig:sankey-2050}; energy flows for cost-optimal results (i.e.\ not minimising or maximising green hydrogen production) are shown in \cref{supfig:sankey-opt-2040,supfig:sankey-opt-2050}.
    }
    \label{fig:sankey}
  \end{adjustwidth}
\end{figure}

Looking at four specific examples of energy flows by 2040 (\cref{fig:sankey}; see \cref{supfig:sankey-2050} for the corresponding energy flows by 2050), we see the substantial direct and indirect impacts of different green hydrogen production pathways on the overall energy system.
Large variations in green hydrogen production induce a difference of up to \qty{3300}{TWh} in emissions-free electricity generation between the scenarios.
Counter-intuitively, maximising green hydrogen production leads to a relative \emph{increase} in natural gas use.
On one hand, reduced use of fossil oil (due to synthetic oil production) allows for a relative increase in natural gas consumption (primarily for heating) within the same emissions and \ch{CO2} storage limits.
Simultaneously, increased green hydrogen production leads to relatively less use of electricity for heating and a slower shift away from natural gas heating.

We see that the future prospects of green hydrogen production, fossil oil and fossil gas are tightly intertwined already in 2040 in systems compatible with the 90\% emissions reduction target.
While a consistent demand for fossil oil remains by 2040, use of fossil gas varies greatly between scenarios with an interquantile range of 50--\qty{900}{TWh} in cost-optimisations (\cref{supfig:fossil-fuels}) --- this rises by an average of \qty{450}{TWh} under green hydrogen maximisations with a cost slack of 5\%.
Large uncertainty around the prospects for green hydrogen creates the risk of stranded assets or missed climate targets --- this risk also extends to the fossil fuel industry as well as CCS facilities.

\subsection*{Sensitivity of green hydrogen production}

\begin{table}
  \centering
  \begin{tabular}{lr@{}l|>{\hspace{2em}}r>{\hspace{2em}}r>{\hspace{2em}}r}
    \multicolumn{6}{r}{\emph{Change in green \ch{H2} production [Mt/a]:}} \\
    \toprule
    \textbf{Setting} & \multicolumn{2}{l|}{$\quad$\textbf{Change}} & \textbf{Min.} (5\%) & \textbf{Opt.} & \textbf{Max.} (5\%) \\
    \midrule
    \begin{filecontents*}{figures/sensitivity-2040-2050.csv}
variable,a,b,min,opt,max
Electrolysis,(a),\ $\rightarrow$ (b),0.5,15.7,28.4
CCS,(a),\ $\rightarrow$ (c),-8.2,-15.1,-20.6
Transportation,(a),\ $\rightarrow$ (c),-7.3,-13.4,-23.6
Biomass,(a),\ $\rightarrow$ (c),-5.2,-8.1,-9.8
Imports,(a),\ $\rightarrow$ (b),-11.3,-8.0,-6.8
Weather year,(a),\ $\rightarrow$ (b),-1.0,0.3,-2.9
\end{filecontents*}
\csvreader[
    head to column names,
    late after line=\\,
    ]
    {figures/sensitivity-2040-2050.csv}{}{\variable & \a & \b & $\min$ & $\opt$ & $\max$} \midrule
                                                                             \begin{filecontents*}{figures/ccs-sensitivity-2040-2050.csv}
variable,a,b,min,opt,max
CO$_2$ seq. potential,275,\ $\rightarrow$ 1100 Mt/a,-20.6,-21.1,-21.3
Carbon capture cost,$+50\%$,\ $\rightarrow$ $-50\%$,0.7,0.2,8.3
CO$_2$ seq. cost,30,\ $\rightarrow$ 15 EUR/t,-0.1,-0.2,-0.6
\end{filecontents*}
\csvreader[
                                                                             head to column names,
                                                                             late after line=\\,
    ]
    {figures/ccs-sensitivity-2040-2050.csv}{}{\emph{\variable} & \emph{\a}& \emph{\b} & $\min$ & $\opt$ & $\max$} \bottomrule
  \end{tabular}
  \caption{
    Sensitivity of green hydrogen production from 2040 to 2050 to scenarios settings (top half) and CCS-related parameters only (bottom half), in Mt/a.
    Higher absolute numbers mean that minimum/optimal/maximum green hydrogen production is more sensitive to the given scenario setting; numbers closer to zero indicate less sensitivity.
    For each of green hydrogen minimisations (``Min.''), cost-optimisations (``Opt.'') and green hydrogen maximisations (``Max''), the sensitivities are calculated as the coefficients in a linear regression predicting green hydrogen production based on scenario settings (Methods).
    For green hydrogen min- and maximisation, we only take results with a 5\% total system cost slack (as indicated in the column headers).
    The linear regression is required as the change from one setting to another is evaluated ``globally'' across all possible combinations of settings for the other categories.
    The numbers can be interpreted as the estimated average change in total green hydrogen production by changing from the first (most pessimistic) to last (most optimistic) settings in each scenario category, across all other scenarios.
    For example, going from (a) pessimistic to (c) optimistic assumptions on CCS is estimated to change total green hydrogen production by \qty{-15.1}{Mt} in cost-optimisations on average across all other scenarios. Green hydrogen production is decreasing since more optimistic assumptions on CCS make fossil fuels a more viable competitor to green hydrogen.
    On the other hand, more optimistic assumptions on electrolysis (going from a 50\% cost increase to a 50\% cost decrease) lead to a positive change in green hydrogen production of \qty{15.7}{Mt} in cost-optimisations on average across all other scenarios.
    Since the CCS settings affect multiple separate parameters (namely, sequestration potential, capital cost of carbon capture and marginal sequestration cost), a separate set of model runs was conducted where these parameters were varied independently (Methods).
    The bottom half of the table is based on this separate sensitivity analysis.
    Note that sensitivities to individual settings are only comparable to a limited degree as they depend on the setting levels.
    For instance, the table shows that going from CCS potential setting (a) to (c) induces a larger change in optimal green hydrogen production than going from electrolyser cost setting (a) and (b), \emph{not} that green hydrogen production is more sensitive to CCS potential than to electrolyser costs in a generalisable sense. 
    See also \cref{supfig:demand-total-h2,supfig:demand-scenario-changes} for an overview over the use of hydrogen broken down by the type of use (e.g. synthetic oil production, ammonia production, etc.), and how this changes between different scenarios.
  }
  \label{tab:sensitivity}
\end{table}

Using a global sensitivity analysis across our scenarios (\cref{tab:sensitivity}), we show that the electrolysis, CCS potential and transportation electrification settings have the biggest impact on cost-optimal levels of green hydrogen production.
A separate analysis shows that out of the different factors making up the CCS potential setting, it is the \ch{CO2} sequestration potential that has the most significant impact by far.

Although a moderate level of green hydrogen production is cost-optimal in most scenarios under consideration (\cref{fig:pathways}), it is possible to eliminate green hydrogen production entirely in more than half of all scenarios while only resulting in a total system cost increase of 2\% or less.
Within our scenario assumptions, minimum green hydrogen production levels are most sensitive to the availability of green fuel imports and of CCS --- the main alternatives to green hydrogen production (\cref{tab:sensitivity}); transportation electrification is also a significant factor.

Various combinations of scenario settings can conspire to make green hydrogen production indispensable; all scenarios where minimised green hydrogen production is non-zero have restricted green imports.
For example, a \emph{combination} of restricted green fuel imports, baseline transportation electrification and pessimistic CCS potential makes green hydrogen production absolutely necessary; in this case minimum green hydrogen production is at least 8.5--\qty{10.0}{Mt} throughout 2040--2050 even allowing a 10\% total system cost increase (see \cref{supfig:CCS-imports}).

\subsection*{Robust corridors of green hydrogen production}

\begin{figure}
  \centering
  \begin{adjustwidth}{-1.5cm}{-1.5cm}
  \includegraphics[width=17cm]{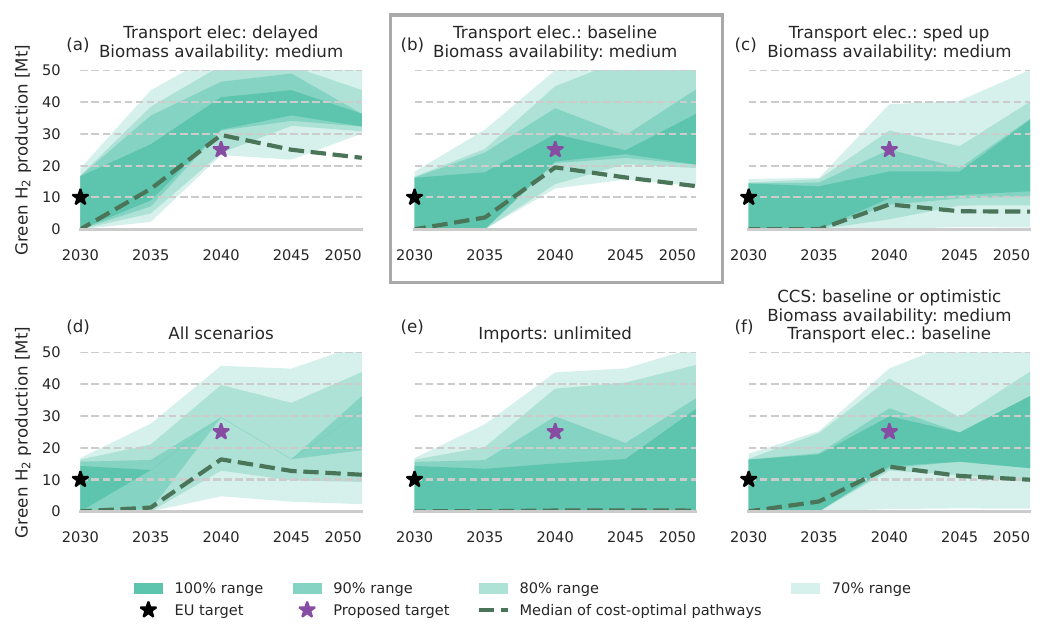}
    \caption{
      Corridors of European green hydrogen production that are feasible and near-optimal (10\% total system slack) under various sets of scenarios.
      (Here, ``feasible'' is used to mean technically feasible --- a valid model solution --- as opposed to ``viable'', i.e. possible within social and environmental constraints \cite{brown-bischof-niemz-ea-2018}.)
      The 100\%, 90\%, 80\% and 70\% ranges cover production levels that are feasible and near-optimal for the respective percentage of scenarios.
      Panels (a)--(f) show these ranges for different selections of scenarios, with panel (d) including all 216 scenarios. 
      For example, panel (b) shows that \qty{25}{Mt} of green hydrogen production is near-optimal and feasible for all scenarios ($216 / (3 \cdot 3) = 24$) with the baseline transportation setting and medium biomass setting; mean while, \qty{40}{Mt} is only near-optimal and feasible for 80\% of those scenarios.
      For a more detailed explanation and exact definitions, see the corresponding methods section.
      Note that around modelling region includes non-EU countries (see also \cref{fig:methodology}), meaning that a robust target for the EU could be lower --- the EU represents about 83\% of the final energy demand in our entire modelling region. 
      Excluding sets of scenarios from consideration expands the robust corridors, because there are fewer scenarios for which any given green hydrogen production level needs to be feasible and near-optimal.
      In panels (d) and (f), there are some time horizons where no single green hydrogen production level is feasible and near-optimal in at least 100\% or 90\% of the scenarios; this is indicated by the respective percentage ranges tapering to width 0; the tapering point is chosen to be the point that is feasible and near-optimal for most scenarios.
      Alternative versions of this figure are given in the Supplementary Information, at total system cost slack levels of 2\% and 5\% (\cref{supfig:robustness-slack-levels}).
      While we focus on scenarios with medium biomass availability in this figure (in panels (a)--(c)), \cref{supfig:robustness-inc-biomass} shows robust corridors under sets of scenarios with low and high biomass availability for comparison.
    }
    \label{fig:robustness}
  \end{adjustwidth}
\end{figure}

Despite the significant sensitivity of green hydrogen to developments in electrolysers, CCS, green fuel imports and more, some green hydrogen production targets remain adequate under most or all considered scenarios.
\cref{fig:robustness} presents robust corridors of green hydrogen production levels (Methods), considering all 216 scenarios as well as various subsets of scenarios.
Our results show that \qty{25}{Mt} of green hydrogen production in 2040 (or indeed, any level approximately between 21 and \qty{30}{Mt}) would be feasible, and the total system cost within 10\% of cost-optimal, for all scenarios with baseline transportation electrification and medium biomass availability (\cref{fig:robustness}, panel b).
The target is also within 5\% of cost-optimal for most (80\%) of the considered scenarios, but not all (\cref{supfig:robustness-slack-levels}).
If all scenarios are considered (including those with low biomass availability and delayed transportation electrification), the \qty{25}{Mt} target is feasible and near-optimal (at 10\% slack) against only 80\% of scenarios (\cref{fig:robustness}, panel d).
In the remaining 20\% of scenarios, producing \qty{25}{Mt} of green hydrogen could either lead to severe cost overruns, or on the other hand could be insufficient to reach the 90\% emissions reduction target.

Notably, the robust corridor lays significantly \emph{above} the median of cost-optimal pathways across all scenarios, which only reaches about \qty{16}{Mt} in 2040.
A robust target hedges against the possibility of limited green fuel imports and low CCS potential.
If CCS and/or green fuel imports \emph{do} scale up rapidly by 2040, producing \qty{25}{Mt} of green hydrogen is still a ``low regret'' option in the sense that total system costs will increase by no more than 5\% in most (80\%) of the considered scenarios (those with baseline transportation and biomass settings; \cref{fig:robustness} panel b), and no more than 10\% in the worst case relative to a cost-optimal trajectory.
In absolute terms, total system cost averages \qty{910}{bn~EUR} in 2040 across all scenarios.

In scenarios where \qty{25}{Mt} is significantly above the cost-optimal production level, reaching the target would require financial incentives; the target being \emph{robust} means that these incentives will raise total system cost by at most 10\%.
Policy-makers could employ subsidies (paid per kilogram of green hydrogen produced) in order to boost production. 
Using results from green hydrogen maximisations (Methods), we estimate that, on average across all scenarios, a subsidy level of \qty{0.55}{EUR/kg\ch{H2}} is required in 2040 to reach the \qty{25}{Mt} target for a total subsidy volume of \qty{13.7}{bn~EUR}.
This should be compared to overall hydrogen prices varying between about 2 and \qty{4}{EUR/kg\ch{H2}} in cost-optimal model runs (\cref{supfig:h2price}).
The highest required subsidy observed across all scenarios is \qty{46.7}{bn~EUR/a}.
The exact subsidy level would have to be adjusted over time to adapt to changing circumstances.

In green hydrogen maximisations, ``excess'' subsidised green hydrogen is not wasted, but consistently used to produced synthetic fuel (\cref{fig:sankey}).
See \cref{supfig:demand-total-h2} for a complete overview over what hydrogen is used for in cost-optimisation as well as green hydrogen min- and maximisations, and \cref{supfig:demand-scenario-changes} for how hydrogen demand is affected by the various scenario settings.
While we focus on green hydrogen here, a target for synthetic fuel production could be a viable policy alternative or complement to a green hydrogen target.

\Cref{fig:robustness} shows that the robust corridor is sensitive to the selection of scenarios under consideration.
If accelerated transportation electrification is assumed in addition to medium biomass availability (panel c), 10--\qty{18}{Mt} of green hydrogen production would be robust, making the \qty{25}{Mt} target too high.
If, on the other hand, transportation electrification is delayed, then the robust range shifts to 31--\qty{41}{Mt}.
As such, only a 5-year delay or speed-up in land transportation electrification can shift a green hydrogen production target for 2040 by about \qty{10}{Mt}.

If pessimistic CCS or green fuel import scenarios are excluded, lower green hydrogen targets could also be robust.
In the case of CCS, the \ch{CO2} sequestration potential is particularly important, with the difference between a maximum injection capacity by 2040 of \qty{125}{Mt\ch{CO2}/a} (pessimistic CCS scenario) or \qty{250}{Mt\ch{CO2}/a} (moderate CCS scenario) explaining most of the difference between panel (d) and (f) (\cref{tab:sensitivity}).
Thus, understanding the feasibility of scaling up geological storage of \ch{CO2} as well as the availability of green fuel imports is especially important when developing robust targets for green hydrogen production.

In a separate analysis (Methods), we also show that any green hydrogen target is highly dependent on the assumed rate of emission reductions.
If a reduction of only 80\% is required by 2040 (as opposed to the baseline of 90\%), no green hydrogen target may be necessary at all (\cref{supfig:robustness-emission-reductions}).
If decarbonisation is accelerated with a 95\% emission reduction target for 2040, a green hydrogen target of \qty{25}{Mt} is insufficient --- a \qty{40}{Mt} target would be adequate.

\subsection*{Limitations}

This study focusses primarily on the supply and conversion side of the European energy system while fixing most final energy demand exogenously (Methods).
While we do vary the single largest exogenous demand for oil in our transportation scenarios, smaller fixed demands are left unexplored in this study. 

Another limitation in this study is the incomplete representation of technological learning; instead, costs of some technologies are assumed exogenously to fall in line with expected estimates.
In reality, heavier initial investment in e.g.\ electrolysers could drive down costs faster.
Imports are represented in limited detail; in particular, different import origins, seasonally varying prices and imports of other energy-related goods apart from fossil- and green fuels are not considered.

Modelling with myopic foresight over the planning horizons (as opposed to perfect foresight) can lead to sub-optimal transition pathways.
In the context of this study this could include stranded assets, --- in particular, electrolysers (which generally have lower capacity factors in 2050 than in 2040; see \cref{supfig:h2-vs-elec-cap}). 

We explicitly consider limitations to \ch{CO2} sequestration as well as green energy imports, but do not implement scaling constraints for other technologies, including renewable energy, electrolysers and carbon capture technology.
Some scenarios (especially maximisations of green hydrogen production) do lead to sharp increases in installed capacities (see e.g. \cref{supfig:elec-cap,supfig:DAC}); such sudden expansion may be impossible or require large-scale market intervention.
The lack of scaling constraints could especially affect our results as they pertain to the near-term (2025--2035).

Our modelling is based on one ``easy'' and one ``difficult'' weather year (2020 and 1987, respectively), but our results are not explicitly ensured to be robust against extreme weather, climate change or other shocks.
\ch{CO2} emissions within Europe are fully represented, but we do not perform any life-cycle analysis on the materials required for the here-in outlined energy system transitions.

\subsection*{Discussion \& Conclusions}

The EU target of \qty{10}{Mt} of green hydrogen production by 2030, while technically feasible, has been criticised as unrealistic and is unlikely to be met; it has been described as a purely political target \cite{europeancourtofauditors-2024}.
Going beyond 2030, a roadmap based on robust analysis is called for.
However, the research literature contains projections ranging from 0 to \qty{160}{Mt} green hydrogen production by 2050 (\cref{suptab:green-h2-lit-review}).
The European energy system could be powered by almost exclusively renewable power and green hydrogen by 2050, or keep relying on significant fractions of abated fossil fuels.
We show that these different directions, and different roles for green hydrogen, can be realised at surprisingly similar total system cost.
This similarity, however, betrays the far-reaching differences in energy flows, with renewables, the continued use of natural gas, the heating sector and synthetic oil production being particularly volatile.
Divergence between radically different pathways already happens in 2040, driven by the ambitious 90\% emissions reduction target.
Given the lead time of over-turning energy infrastructure, this uncertainty presents a significant challenge to the energy transition.

We make the case that the Europe risks little by committing to producing some \qty{25}{Mt} of green hydrogen by 2040, helping reduce investment risk; such a target also seems likely to be technologically feasible \cite{odenweller-ueckerdt-ea-2022}.
If founded in a thorough feasibility analysis and presented alongside plausible implementation mechanisms, such a target could stand a better chance to be met than the 2030 target.
With the EU representing approximately 83\% of final energy demand in our modelling region (which also includes the UK, Switzerland, Norway and most Balkan countries), an EU target of \qty{20}{Mt} could be considered.
Such a target hedges against the risk of delays and unavailability of CCS and green fuel imports.

If land transportation electrification is accelerated by 5 years, the target could be adjusted down by \qty{10}{Mt}.
Delayed electrification, in addition to leading to significantly higher overall system costs \cite{zeyen-kalweit-ea-2025}, would require a higher green hydrogen production target.
In this case, our results indicate that significant use of synthetic fuel in the land transportation sector may be the most effective (or indeed only) way of reaching the 90\% emissions reduction target; a controversial prospect given that direct electrification is much more energy-efficient.
While ending the sale of new \ch{CO2}-emitting cars has already been delayed from 2030 to 2035 \cite{europeancommission-2022a}, our results show the importance of preventing further such delays.
While financial incentives or other policies \cite{odenweller-ueckerdt-2024} may be needed in most scenarios to reach the \qty{25}{Mt} target (with the median cost-optimal pathway across all scenarios having only around \qty{16}{Mt} of green hydrogen production in 2040), our results show that total system cost increases by no more than 5\% over cost-optimal in most (80\%) scenarios, and no more than 10\% in any scenario (assuming medium, current-day biomass availability and baseline transportation electrification).
We estimate that a subsidy of \qty{0.55}{EUR/kg\ch{H2}} would be required on average, for a total of \qty{13.7}{bn~EUR/a}.

Green hydrogen production based on local electricity has the additional benefit of increasing European energy independence.
Moreover, the alternative of relying on fossil fuels combined with \ch{CO2} sequestration is fundamentally unsustainable in the long run.
Not only are fossil fuel reserves depleted over time, but the capacity for geological storage of \ch{CO2} in Europe is also finite (conservatively estimated to \qty{115}{Gt} \cite{anthonsen-christensen-2021}) and could run 
out within a century under more intensive use of CCS.

We stress that even in scenarios with little geological storage of \ch{CO2} (CCS), we still see large-scale carbon capture and utilisation (CCU) --- especially when maximising green hydrogen production.
The vast majority of cost-optimal pathways exhibit annual carbon capture volumes of more than \qty{350}{Mt\ch{CO2}} by 2040 (\cref{supfig:captured-co2}), necessitating faster growth than those observed historically in wind, solar and nuclear power \cite{kazlou-cherp-ea-2024}.
In some scenarios, significant use of direct air capture of \ch{CO2} becomes cost-optimal already in 2040 (\cref{supfig:DAC}).
European targets for scaling up carbon capture, regardless of whether for storage or utilisation, should be considered.

Green hydrogen comes with its own risks in terms of difficulty in scaling up \cite{odenweller-ueckerdt-ea-2022} and the global warming potential of leaked hydrogen \cite{sand-skeie-ea-2023}.
Moreover, a green hydrogen-based energy system requires vast quantities of renewable energy, which can lead to land use conflict and social acceptance issues.
Falling demand for liquid fuels from 2040 to 2050 could also risk leaving hydrogen electrolysers a partially stranded asset by 2050, with falling capacity factors (\cref{supfig:elec-cap}) and prices (\cref{supfig:h2price}).

However, the continued use of fossil fuels combined with CCS is unsustainable, and a temporary solution at best.
Meanwhile, Europe is trying to reduce its dependence on energy imports after the 2022 gas crisis, and the future development of global green fuel markets is highly uncertain.
Improving our understanding of how quickly \ch{CO2} storage and green fuel imports can scale up at the minimum is crucial for setting the best possible green hydrogen target.
Developing a clear green hydrogen strategy for 2040 and beyond will improve the chances for Europe to reach its ambitious climate targets.

\subsection*{Code and data availability}

The code to reproduce the results of the present study, as well as links to the data used, are available at \url{https://github.com/koen-vg/eu-hydrogen/tree/v1.1}.
All code is open source (licensed under GPL v3.0 and MIT), and all data used are open (various licenses).

\subsection*{Acknowledgements}

ERA5 reanalysis data \cite{hersbach-bell-ea-2018} were downloaded from the Copernicus Climate Change Service (C3S).

The results contain modified Copernicus Climate Change Service information 2020. Neither the European Commission nor ECMWF is responsible for any use that may be made of the Copernicus information or data it contains.

\subsection*{Declaration of interests}

The authors declare no competing interests.

\small
\subsection*{Methods}

\subsubsection*{Scenario assumptions}

Our scenario assumptions consist of all possible combinations of the following levels in five different settings:
\begin{enumerate}
\item CCS potential:\\
  \textbf{(a)} \ch{CO2} sequestration potential of 25/125/\qty{275}{Mt/a} by 2030/2040/2050, cost of sequestration of 30 EUR/t\ch{CO2}, capital cost of carbon capture $+50\%$.\\
  \textbf{(b)} \ch{CO2} sequestration potential of 50/250/\qty{550}{Mt/a} by 2030/2040/2050, cost of sequestration of 20 EUR/t\ch{CO2}.\\
  \textbf{(c)}  \ch{CO2} sequestration potential of 100/500/\qty{1100}{Mt/a} by 2030/2040/2050, cost of sequestration of 15 EUR/t\ch{CO2}, capital cost of carbon capture $-50\%$.
\item Biomass potential: following ENSPRESO \cite{ruiz-nijs-ea-2019} scenarios:\\
  \textbf{(a)} Low (\qty{450}{TWh} solid biomass, \qty{180}{TWh} biogas by 2050). \\
  \textbf{(b)} Medium (\qty{1190}{TWh} solid biomass, \qty{350}{TWh} biogas by 2050).\\
  \textbf{(c)} High (\qty{2830}{TWh} solid biomass, \qty{530}{TWh} biogas by 2050).
\item Green fuel imports:\\
  \textbf{(a)} Total imports cannot exceed domestic green hydrogen production.\\
  \textbf{(b)} Unrestricted imports.
\item Electrolyser capital costs: \textbf{(a)} $+50\%$. \textbf{(b)} $-50\%$.
\item Land transportation electrification:\\
  \textbf{(a)} Delayed by 5 years with respect to baseline.\\
  \textbf{(b)} Baseline.\\
  \textbf{(c)} Sped up by 5 years with respect to baseline.
\item Weather year: \textbf{(a)} 1987. \textbf{(b)} 2020.
\end{enumerate}

In the above, \ch{CO2} sequestration potential is linearly interpolated for 2035 and 2045; it is set to 0 in 2025 for all scenarios.
The middle-of-the-road trajectory of 50/250/550 Mt/a of \ch{CO2} sequestration by 2030/2040/2050 reflects the official target for 2030 set out in the EU Net Zero Industry Act \cite{eu-net-zero-industry-2024}, as well as modelled EU estimates for 2040.
In particular, the carbon sequestration limit of \qty{250}{Mt/a} by 2040 is primarily based on the EU 2040 climate target impact assessment \cite{europeancommission-2024}, part 3, section 1.1.3.2.
The impact assessment uses four different scenarios and is based in part on modelling with PRIMES; the different scenarios indicate carbon sequestration between 90 and \qty{350}{Mt/a} by 2040.
The three out of four scenarios achieving a 90\% emissions reduction by 2040 sequester more than \qty{200}{Mt/a} of \ch{CO2} by 2040.
Several other sources are cited in the impact assessment which point to comparable levels of \ch{CO2} sequestration needed by 2040.
For 2050, the Net Zero Industry Act \cite{eu-net-zero-industry-2024} of EU Commission mentions internal research indicating that \qty{550}{Mt/a} may be necessary to achieve net zero emissions; we use this as middle-of-the-road baseline.
It is comparable to the \qty{500}{Mt/a} feasibility limited used by the European Environmental Agency \cite{europeanenvironmentagency-2023a}.
For our pessimistic and optimistic CCS setings, we half and double the sequestration limit, respectively.

The literature on cost estimates for \ch{CO2} sequestration (including transportation and permanent storage) is sparse.
Smith et al.~\cite{smith-morris-ea-2021a}, while noting that many integrated assessment models have traditionally assumed a cost of \qty{10}{USD/t}, finds a range of costs, from 4 to \qty{45}{USD/t}.
Adjusting for recent inflation, we assume \qty{15}{EUR/t} as a baseline cost comparable to the traditional \qty{10}{USD/t}.

By ``capital cost of carbon capture'', we denote the capital cost premium of all components that capture \ch{CO2}, including blue hydrogen production (steam methane reformation with carbon capture), combined heat- and power plants with carbon capture, industrial processes with carbon capture and direct air capture of \ch{CO2}.
We apply multipliers of $-50\%$ and $+50\%$ in the optimistic and pessimistic settings, respectively.
For components which have versions with and without carbon capture, the multiplier is applied to the difference between the capital costs of the versions with and without carbon capture.
For direct air capture (DAC) of \ch{CO2}, the multiplier is applied to the entire capital cost.

Biomass potentials over time in the low, medium and high settings are shown in \cref{tab:biomass-potentials}.

\begin{table}
  \centering
  \begin{tabular}{rllllll}
    \toprule
    Year & \multicolumn{3}{c}{Solid biomass [TWh]} & \multicolumn{3}{c}{Biogas [TWh]} \\ \cmidrule(lr){2-4} \cmidrule(lr){5-7} & (a) Low & (b) Medium & (c) High & (a) Low & (b) Medium & (c) High \\\midrule
    \begin{filecontents*}{figures/biomass_potentials.csv}
2025,217,382,869,56,113,170
2030,366,688,1733,114,228,342
2035,535,1027,2574,174,348,522
2040,521,1020,2541,177,351,525
2045,427,1017,2585,179,354,529
2050,336,1014,2628,181,358,533
\end{filecontents*}
\csvreader[
    no head,
    late after line=\\,
    ]{figures/biomass_potentials.csv}{}
    {\csvcoli & \csvcolii & \csvcoliii & \csvcoliv & \csvcolv & \csvcolvi & \csvcolvii} \bottomrule
  \end{tabular}
  \caption{Biomass (solid and gas) potential over time in low (a), medium (b) and high (c) settings, used in scenario assumptions. Values are in TWh/a. The potentials are extracted from the ENSPRESO database \cite{ruiz-nijs-ea-2019} and include only sustainable biomass potential (hence the low value in initial time horizons).}
  \label{tab:biomass-potentials}
\end{table}

Green fuel imports include energy imports in the form of shipped liquid green hydrogen, ammonia, methanol, synthetic oil and synthetic gas, with fixed costs (though decreasing by planning horizon) taken from Hampp et al.~\cite{hampp-duren-ea-2023}; see \cref{tab:green-fuel-import-costs}.
The limit applied to total green fuel imports in the pessimistic setting (a) is not fixed, but rather limits imports by the amount of European green hydrogen production in terms of energy content (using the lower heating value).
This can also be seen clearly in \cref{supfig:h2-vs-imports}.

Baseline electrolyser costs are unchanged from PyPSA-Eur, and are derived from the \texttt{technology-data} repository.
They fall over time; upfront investment costs are show in \cref{tab:electrolyser-costs} (also displaying efficiencies over time).
In the model these costs (like all other capital costs) are annualised.

\begin{table}
  \centering
  \begin{tabular}{rllllll}
    \toprule
    Year & 2025 & 2030 & 2035 & 2040 & 2045 & 2050 \\ \midrule
    Capital cost [EUR/kW] & 2152 & 1793 & 1614 & 1435 & 1315 & 1196 \\
    Efficiency [\%] & 58.7 & 62.2 & 63.7 & 65.3 & 67.6 & 69.9 \\ \bottomrule
  \end{tabular}
  \caption{Baseline electrolyser capital costs and efficiency over time. Efficiency is in terms of the lower heating value of the produced hydrogen.}
  \label{tab:electrolyser-costs}
\end{table}

The land transportation electrification setting affects the speed of the transition to battery electric (and fuel cell electric) light and heavy vehicles; see below for more details on the assumed baseline transition pathway.
In settings (a) and (c), the transition pathway is simply shifted backwards and forwards by 5 years, respectively. In (a), the pessimistic setting, 2025 fuel shares are kept constant until 2030, and only start transitioning from 2035 and onwards. Similarly, in (c), the optimistic setting, fuel shares are the same in 2025 as in the baseline transition pathway, but then jump forward by 5 extra years so that the assumed fuel mix in 2030 is equal to the fuel mix in 2035 in the baseline transition, and so forth.
See also \cref{tab:transport-fuel-shares} for the exact fuel share assumptions.

The selected weather years represent a relatively difficult (1987) and easy (2020) year from the 1980--2020 period \cite{grochowicz-vangreevenbroek-ea-2023,grochowicz-vangreevenbroek-ea-2024}.

The 216 scenarios, 6 planning horizons and cost-optimisations as well as green hydrogen min- and maximisations with 3 different slack levels make for a total of 1296 cost-optimisations and 3888 min- and maximisations each.
Some of the green hydrogen maximising model runs fail due to timing out or numerical stability issues.
In total, 1 / 648 maximisations fail at the 2040 horizon, 250 / 648 fail at the 2045 horizon and 271 / 648 fail at the 2050 horizon.
A single green hydrogen minimisation fails at the 2045 and 2050 time horizons.
In summary, the upper ranges of green hydrogen production at the 2045 and 2050 time horizons are only partially covered by our results.

\subsubsection*{Model set-up}

We use the capacity expansion model PyPSA-Eur \cite{PyPSAEur,PyPSAEurSec} (version 0.13 with minor modifications) to generate feasible energy system designs at planning horizons $Y=2025,2030,2035,2040,2045,2050$.
The model includes representations of the electricity, heating, transportation and industrial sectors, maintaining energy balances across electricity, hydrogen, gas, oil, biomass, ammonia and methanol as well as explicitly keeping track of \ch{CO2} balances.
Each energy carrier may be produced and consumed by various processes; ``gas'' could be obtained as natural gas or synthesised from hydrogen, for example.
Our modelling region consists of all EU countries except Cyprus and Malta, while also including Albania, Montenegro, North Macedonia, Norway, Serbia, Switzerland and the United Kingdom.

The model is run with myopic planning foresight: optimisations are carried out at each planning horizon without foresight to future time horizons.
At each planning horizon, those capacities from previous horizons that have not yet reached the end of their lifetime are carried over.
Currently existing power plants, transmission infrastructure, gas pipelines and storage and heating infrastructure are added to the initial 2025 model.
Costs are given in 2023 EUR; investment costs are annualised with a discount rate of 7\%.
Technology costs and parameters are retrieved from the \texttt{technology-data} repository\footnote{\url{https://github.com/PyPSA/technology-data}}, version 0.9.2, unless otherwise indicated.

The following exposition serves as a brief summary of the most relevant features of the model.
Unless otherwise specified, the summary mainly touches on existing model features and assumptions; the official documentation\footnote{\url{https://pypsa-eur.readthedocs.io/en/stable/}} serves as a more exhaustive reference and contains sources to all data used.
The supplementary information to a recent publication by Neumann et al. \cite{neumann-zeyen-ea-2023} also contains a number of useful figures documenting basic behaviour of the model.
In the below outline, we clearly indicate changes made to PyPSA-Eur for the purpose of this study.

After covering energy demand and supply in the model generally, we summarise demand and supply of hydrogen specifically for clarity.

\subsubsection*{Energy demand}

Electricity demand profiles in PyPSA-Eur are fixed exogenously and based on current demand, but electrified heating demand is subtracted.
Heating demand (a combination of space heating and hot water demand) is also fixed exogenously but modelled separately from the electricity sector; demand is assumed to decrease gradually towards 2050 to reflect improving building standards.
Heating demand depends on ambient temperatures, and can be satisfied through a combination of combined heat and power plants, heat pumps, resistive heating, gas boilers, biomass boilers, solar thermal collectors, waste heat from industrial processes and long term energy storage.

Other energy demand is largely retrieved from Eurostat at 2019 levels (i.e. the last year before the COVID-19 pandemic) and kept fixed over time.
For the transportation sector, final land transportation, shipping and aviation demand is fixed exogenously to 2019 levels; fuel mixes (i.e. fraction of electric cars, ICE cars, etc.) are also specified exogenously but set to change over time towards 2050.
The assumptions on fuel mixes in the transportation sector are \textbf{changed} from the PyPSA-Eur default to reflect more realistic and up-to-date estimates.
For land transportation, we use projections from the impact assessment accompanying the proposed EU 2040 climate target \cite{europeancommission-2024}, as shown in \cref{tab:transport-fuel-shares}; while the cited impact assessment covers the EU, we extrapolate fuel mix ratios to the whole modelling region of this study. 
The projections indicating a gradually increasing share of electricity consumption to 49.3\% and 71.9\% in the land transportation sector (including both light and heavy-duty vehicles, weighted by energy demand) by 2040 and 2050, respectively.
The share of fuel cell vehicles is projected to increase to 18\% by 2050; the share of ICE vehicles falling to 10.1\% by 2050.

While more than 60\% of passenger cars are assumed to be electrified by 2040, this is offset by a slower electrification of heavy vehicles \cite{europeancommission-2024}, resulting in an aggregate electrification rate (in terms of energy consumption) of just less than 50\%.
The same electrification rate for 2040 is predicted in a recent report by PwC and the Fraunhofer institute \cite{neuhausen-rose-ea-2023}, which also covers the EU and predicts sales of new ICE vehicles to drop to zero in 2035 for passenger cars and in 2040 for heavy vehicles.
On the other hand, market penetration for electric passenger cars reaching 60\% ``appears to be very optimistic given
fleet turnover dynamics of passenger cars in Europe'' according to another study looking at the contribution of the transportation sector towards to 2040 climate target \cite{seibert-kasten-ea-2024}.
In this study, it is pointed out that the average lifespan of passenger cars is significantly longer in Eastern European states, and that used car imports contribute more to fleet turnover than new vehicle registration, potentially slowing down European passenger car fleet turnover.

For the shipping sector, we \textbf{add} the capacity to use natural gas and ammonia as a shipping fuel (in addition to oil, methanol and hydrogen which are implemented by default), and \textbf{change} the assumed fuel mix development in the shipping sector in line with a study \cite{transportenvironment-2023} on the modelled impact of the recent FuelEU Maritime regulation \cite{fueleu-2023} (Scenario 1A in the study); see \cref{tab:transport-maritime-shipping-shares}.
The use of fossil oil in the shipping sector is set to decrease to 0 by 2050, replaced gradually by natural gas (LNG), reaching a share of 40\% by 2035, and thereafter increasing shares of ammonia, reaching 59\% by 2050.
For both of the above, we use relative fuel shares compiled in EU studies even though the modelling region of this study includes additional countries.
The aviation sector is assumed to maintain a constant demand for kerosene; accounted in the model under the ``oil'' resource.

\begin{table}[tb]
  \centering
  \small
  \begin{minipage}{0.3\linewidth}
    \centering
    \textbf{(a)} Delayed by 5 years
    \begin{tabular}{rlll}
      \toprule
      Year & ICE & FC & BEV \\ \midrule 
      2025 & 0.938 & 0     & 0.062 \\
      2030 & 0.938 & 0     & 0.062 \\
      2035 & 0.873 & 0.004 & 0.123 \\
      2040 & 0.511 & 0.039 & 0.45  \\
      2045 & 0.434 & 0.073 & 0.493 \\
      2050 & 0.267 & 0.127 & 0.606 \\
      \bottomrule
    \end{tabular}  
  \end{minipage}
  \hspace{0.5cm}
  \begin{minipage}{0.3\linewidth}
    \centering
    \textbf{(b)} Baseline
    \begin{tabular}{rlll}
      \toprule
      Year & ICE & FC & BEV \\ \midrule 
      2025 & 0.938 & 0     & 0.062 \\
      2030 & 0.873 & 0.004 & 0.123 \\
      2035 & 0.511 & 0.039 & 0.45  \\
      2040 & 0.434 & 0.073 & 0.493 \\
      2045 & 0.267 & 0.127 & 0.606 \\
      2050 & 0.101 & 0.180 & 0.719 \\
      \bottomrule
    \end{tabular}  
  \end{minipage}
  \hspace{0.5cm}
  \begin{minipage}{0.3\linewidth}
    \centering
    \textbf{(c)} Sped up by 5 years
    \begin{tabular}{rlll}
      \toprule
      Year & ICE & FC & BEV \\ \midrule 
      2025 & 0.938 & 0     & 0.062 \\
      2030 & 0.511 & 0.039 & 0.45  \\
      2035 & 0.434 & 0.073 & 0.493 \\
      2040 & 0.267 & 0.127 & 0.606 \\
      2045 & 0.101 & 0.180 & 0.719 \\
      2050 & 0.101 & 0.180 & 0.719 \\
      \bottomrule
    \end{tabular}  
  \end{minipage}
  \caption{Assumed shares of internal combustion engine (ICE), fuel cell (FC) and battery electric (BEV) vehicles by modelling horizon, retrieved from an impact study on the EU 2040 climate target \cite{europeancommission-2024}. The shares represent fractions of the total vehicle stock (including light and heavy-duty vehicles, weighted by energy demand) at each time horizon, not the number of new vehicles sold.}
  \label{tab:transport-fuel-shares}
\end{table}

\begin{table}[tb]
  \centering
  \small
  \begin{tabular}{rllll}
    \toprule
    Year & Oil & Gas & Ammonia & Methanol \\ \midrule
    2025 & 0.85 & 0.11 & 0    & 0.04 \\    
    2030 & 0.69 & 0.27 & 0    & 0.04 \\
    2035 & 0.50 & 0.40 & 0.06 & 0.04 \\
    2040 & 0.33 & 0.39 & 0.24 & 0.04 \\
    2045 & 0.17 & 0.37 & 0.42 & 0.04 \\
    2050 & 0    & 0.37 & 0.59 & 0.04 \\
    \bottomrule
  \end{tabular}
  \caption{Assumed shares of fuel used in maritime shipping for each modelling horizon, retrieved from a modelling study by Transport \& Environment on the impacts of the FuelEU Maritime legislation \cite{transportenvironment-2023}. The study anticipates lower costs for ammonia than for methanol in the long term, but a slow uptake due to the low technological readiness of ammonia as a shipping fuel. Alternative scenarios are also explored in said study; we only consider the baseline scenario for computational reasons and since both ammonia and methanol constitute comparable potential demands for hydrogen.}
  \label{tab:transport-maritime-shipping-shares}
\end{table}

The industrial sector contributes with exogenously specified (but partially changing over time) demand for a number of different energy carriers.
Industrial electricity demand is accounted for in the electricity sector.
However, the steel industry initially induces a demand for coal (for use in blast furnaces); this is assumed to gradually switch to a direct reduced iron process using hydrogen.
With coal being fully phased out, the steel industry induced a demand for hydrogen of about \qty{79}{TWh} by 2050.
The fraction of primary steel is also assumed to decrease from 60\% to 30\% by 2050.
A constant demand for \qty{16.5}{Mt/a} of ammonia (equivalent to \qty{85.4}{TWh} in lower heating value) is assumed for use in the production of fertilizer.
There is also a minor amount of constant demand for methanol (equivalent to \qty{8.7}{TWh/a} in lower heating value) from the chemical industry.
The chemical industry contributes with a significant baseline demand for oil, from \qty{918.7}{TWh/a} in 2025 to \qty{276.8}{TWh/a} in 2050 (declining due to an assumed decline in plastics demand and increase in recycling fraction).
While not strictly speaking part of industry sector, the model also assumes a constant \qty{102.8}{TWh/a} oil demand for the agriculture sector.
All of the above follows PyPSA-Eur defaults; data is primarily sourced from Eurostat.

Handling of process heat demand in industry is \textbf{changed}: while PyPSA-Eur, as of version 0.13, exogenously specifies demand for natural gas and solid biomass in order to serve industrial process heat, we adapt PyPSA-Eur to allow for endogenous optimisation of fuel selection.
This adaptation is based on a pull request\footnote{\url{https://github.com/PyPSA/pypsa-eur/pull/611}}, open at the time of writing; process heat demand is split in low-, medium- and high-temperature segments.
A constant demand is given for each of these segments; low-temperature heat can be sourced from solid biomass, heat pumps, electric boilers and gas, medium-temperature from solid biomass, gas and hydrogen, and high-temperature heat from gas and hydrogen.
Recall that ``gas'' is treated as an energy carrier that may be sourced as natural gas, biogas or synthesised from hydrogen.

\Cref{supfig:demand-fixed-major} shows exogenously fixed demand for electricity and oil, and \cref{supfig:demand-fixed-minor} shows exogenously fixed demand for gas, hydrogen, ammonia and methanol.

\subsubsection*{Energy supply, conversion and storage}

Electricity can be generated by a variety of different power plants.
Expandable variable renewable generation includes solar PV, onshore wind, bottom-fixed offshore wind and floating offshore wind --- currently existing capacities are also added to the initial 2025 model.
Nuclear power plants are included at their currently existing capacities; nuclear is also expandable.
Existing hydro power plants are also included, but not expandable.
Moreover, existing thermal power plants running on natural gas, oil, coal and biomass are included; gas power plants as well as Allam-cycle methanol power plants are expandable.
The latter thermal power plants consume fuel in order to generate electricity.

Note that no scaling constraints are implemented on any expandable energy supply, conversion or storage technology (except for green energy imports as per the scenario assumptions), meaning that each expandable component (e.g. renewable energy, electrolysis) can be sized without constraint at each time horizon. 

Fossil gas and oil as well as coal are available to the model at market prices; the gas price is assumed to be a constant \qty{29.4}{EUR/MWh} and the oil price \qty{63.3}{EUR/MWh} as per PyPSA-Eur defaults.

Solid biomass and biogas are also available at market prices (\qty{16.3}{EUR/MWh} and \qty{74.6}{EUR/MWh} respectively, with biogas incurring additional costs for upgrading) but only in limited quantities which depend on biomass scenario setting.
We use low, medium and high availability projections from the ENSPRESO \cite{ruiz-nijs-ea-2019} database; PyPSA-Eur uses the medium scenario by default.
Only waste and residual products are considered for biomass, no crops grown specifically for the purpose of bio energy generation.

We \textbf{change} PyPSA-Eur to allow for the import of gas, oil, ammonia, methanol and hydrogen as ``green'' fuels, meaning that the fuels are carbon-neutral.
No carbon emissions can be associated to the production of these fuels, and any carbon contained in the fuel (in the case of gas, oil and methanol) must be sourced directly or indirectly from the atmosphere, not from fossil sources.
Import costs are taken from Hampp et al.~\cite{hampp-duren-ea-2023} and shown here in \cref{tab:green-fuel-import-costs}.
For simplicity, fuels which are spatially resolved (namely, gas and hydrogen) can be imported to any node in the model.

\begin{table}[tb]
  \centering
  \small
  \begin{filecontents*}{figures/import_costs.csv}
Carrier,2025,2030,2035,2040,2045,2050
Hydrogen,122.9,122.9,112.0,101.2,92.7,84.1
Ammonia,120.1,120.1,110.3,100.6,94.5,88.4
Methanol,165.6,165.6,151.9,138.2,128.5,118.7
Oil,234.2,234.2,214.0,193.9,178.6,163.4
Gas,128.4,128.4,118.2,108.1,102.2,96.2
\end{filecontents*}
\csvautobooktabular{figures/import_costs.csv}
  \caption{Assumed green fuel import costs in EUR/MWh (using the lower heating value) from Hampp et al.~\cite{hampp-duren-ea-2023} by energy carrier and year; the cited study uses the same base repository for technology costs (the \texttt{technology-data}  repository: \url{https://github.com/PyPSA/technology-data}), which should alleviate any large discrepancies between foreign and domestic energy production and conversion costs. Processing and transportation costs as well as the cost of sourcing carbon from the atmosphere as feedstock where applicable are included. Transport is assumed to take the form of shipping from the cheapest source; pipeline imports are not included.}
  \label{tab:green-fuel-import-costs}
\end{table}

A number of energy conversion processes are available in the model; each of these is expandable at a cost and subject to optimisation.
Hydrogen can be produced from gas via steam methane reformation with or without carbon capture, from electricity via water electrolysis, from ammonia via ammonia cracking and from solid biomass (with carbon capture).
Ammonia can be produced from hydrogen via the Haber-Bosch process.
Methanol can be produced from hydrogen and \ch{CO2} via methanolisation, as well as from biomass with or without carbon capture.
Oil, apart from fossil sources, can also be produced from hydrogen and \ch{CO2} via the Fischer-Tropsch process as well as from biomass.
Gas, apart from fossil sources, can also be produced from hydrogen and \ch{CO2} via the Sabatier process as well as from biogas (as mentioned above).

For uses of hydrogen, see below.
Methanol is used to satisfy fixed demand from the industry and shipping sectors, and can be used to generate electricity.
Ammonia is also used to satisfy fixed demand from the industry and shipping sectors, and can be cracked to produce hydrogen.

A number of energy storage technologies are also available.
Electricity storage is available directly via expandable grid-connected batteries, as well as non-expendable pumped hydro storage included at today's capacity of about \qty{8.7}{TWh} in total with \qty{56.9}{GW} of dispatch capacity in total.
Existing reservoir hydro is also included with a storage capacity of about \qty{152.4}{TWh} in total with \qty{103.1}{GW} of dispatch capacity in total.
Finally, a growing amount of battery storage from electric vehicles becomes available; the sizing of this storage is not subject to optimisation as it follows the exogenously specified share of electric vehicles.
Under the assumption that 50\% of electric vehicles participate in demand-side management and can provide vehicle-to-grid (V2G) services, the total amount of storage available to the model reaches about \qty{5.2}{TWh} in 2050, with a maximum theoretical dispatch capacity of about \qty{2.3}{TW}.
Electric vehicle battery storage is constrained to be charged to at least 75\% at 7:00 every morning, and the fraction of grid-connected vehicles follows a daily profile with an average of 80\% of all V2G-enabled vehicles being connected to the grid at any one time.

Hydrogen can be stored underground in salt caverns where available \cite{caglayan-weber-ea-2020}, and in compressed tanks where salt caverns are not available, both expandable.
A technical potential for around \qty{2700}{TWh} of underground hydrogen storage exists, only a small fraction of which is ever used by the model (\cref{supfig:h2-storage}).

Gas storage is also possible underground.
About \qty{1300}{TWh} of existing gas storage infrastructure is included; while gas storage is also further expandable, this option is not exploited by the model.
Oil and methanol can both be stored in expandable tanks; ammonia can be liquefied and stored in tanks.
Finally, expandable hot water tanks are available for heat storage.

\subsubsection*{Summary of supply and demand for hydrogen}

Hydrogen can be produced by electrolysis, by steam methane reformation with or without carbon capture, from biomass with carbon capture, by ammonia cracking and can be imported.
Production of hydrogen is endogenously optimised.

Note that, in the context of this study, we refer to any hydrogen produced by electrolysis as ``green hydrogen''; we do not consider restrictions on the carbon intensity of the electricity supply.

An exogenously fixed demand for hydrogen comes from the steel industry for direct iron reduction (\qty{79}{TWh/a} by 2050) as well as the transportation sector (\qty{304}{TWh/a} by 2050) --- see \cref{supfig:demand-fixed-minor}.

Indirect demand for hydrogen also comes from exogenously fixed demand for ammonia (which can only be produced from hydrogen or imported) and methanol (which can be produced from hydrogen, biomass or be imported) --- see \cref{supfig:demand-fixed-minor}.

Hydrogen can be used to produce electricity using fuel cells.
Finally, hydrogen can be used in the model to produce synthetic oil and gas (through the Fischer-Tropsch and Sabatier processes, respectively) as well as to supply medium- and high-temperature industrial process heat and space heating.
By 2040 and onwards, the majority of hydrogen, beyond satisfying a small fixed demand, is typically used in the production of synthetic oil (\cref{fig:sankey}).

\subsubsection*{The carbon cycle and emissions}

\ch{CO2} is tracked explicitly throughout the model.
Through emissions, capture and underground storage, units of \ch{CO2} can be created and moved between the atmosphere, temporary storage (i.e. in overground storage tanks) and permanent storage (underground, e.g. in depleted gas fields).
Any use of carboniferous fuel (oil, gas, methanol), either at final energy demand or in conversion processes, adds units of \ch{CO2} to the atmosphere according to the carbon intensity of the fuel and the nature of the conversion process if applicable.
The use of biomass does not add \ch{CO2} to the atmosphere since the carbon in biomass is considered to originate from the atmosphere.
Where biomass is used as a feedstock to produce carboniferous fuel, \ch{CO2} is subtracted from the atmosphere to compensate for the later burning/use of that fuel.
This is done because the model does not differentiate carboniferous fuel (oil, gas and methanol) by origin; any such fuel is ``mixed'' in the same balancing bus (or buses, in the case of gas, which is balanced by location, not model-wide).
Subtracting \ch{CO2} from the atmosphere in the production of biofuels can also be thought of as analogues to biogenic carbon ultimately being sourced from the atmosphere through photosynthesis.
Likewise, the carbon content of imported green fuels (oil, gas and methanol are the carboniferous fuels that can be imported in ``green'', carbon-neutral versions) is also subtracted from the atmosphere.

Non-energy net emissions are also added to or removed from the atmosphere.
Industrial process emissions that do not directly arise from energy use constitute \qty{190}{Mt/a} in 2025 and gradually reduce to \qty{123}{Mt/a} by 2050; these emissions mainly come from the cement industry, with integrated steel works and the high value chemical industry also playing minor roles.
We \textbf{change} PyPSA-Eur to also include emissions from the agricultural sector on one hand, in \ch{CO2}-equivalents, and negative emissions from the land use, land-use change, and forestry (LULUCF) sector; both are only included for EU countries and sourced from the 2040 EU climate target impact assessment \cite{europeancommission-2024} (using the LIFE scenario).
Combined, agriculture and LULUCF net emissions in the EU start at about \qty{155}{Mt/a} in 2020, and are expected to fall to about \qty{-120}{Mt/a} by 2050.
See \cref{tab:agri-lulucf} for a complete overview over assumed agriculture and LULUCF emissions.

\begin{table}[tb]
  \centering
  \begin{tabular}{lrrrrr}
    \toprule
    Sector & 2015 & 2020 & 2030 & 2040 & 2050 \\ \midrule
    Agriculture & 385 & & 361 & 271 & 269 \\
    LULUCF & & $-230$ & $-310$ & $-360$ & $-389$ \\
    \bottomrule
  \end{tabular}
  \caption{Assumed emissions in Mt of \ch{CO2}-equivalents from the agricultural and LULUCF sectors in the EU. At empty places and 5-year intervals not shown in this table, values are linearly interpolated. The negative emissions of \qty{-310}{Mt/a} by 2030 is an EU target by law \cite{eu-lulucf-2023}; the remaining figures are sourced from the 2040 EU climate target impact assessment \cite{europeancommission-2024}.}
  \label{tab:agri-lulucf}
\end{table}

A number of energy conversion processes as well as final energy demand and process emissions have alternatives with carbon capture (CC) in the model, in which case they deposit a fraction of the \ch{CO2} that would otherwise be emitted to the atmosphere in temporary stores instead.
Technologies which have CC-enabled alternatives are steam methane reformation, combined heat and power plants, gas power plants (by way of Allam-cycle gas power plants), biomass-to-methanol, use of gas or biomass for industrial process heat as well as process emissions (e.g. CC at cement plants).
Direct air capture of \ch{CO2} is also an option, and moves \ch{CO2} from the atmosphere to temporary storage.

Temporarily stored \ch{CO2} can be used (together with hydrogen) as a feedstock for the production of synthetic oil, gas and methanol.
Finally, temporarily stored \ch{CO2} can also be sequestered (stored permanently underground).
Temporary storage of \ch{CO2} is forced to be ``cyclic'' over each optimisation horizon, meaning that the storage levels for \ch{CO2} must be the same at the start and end of the optimisation horizon.

Emissions are limited by restricting the net amount of \ch{CO2} released to the atmosphere, i.e., the difference between release and capture.
\ch{CO2} emissions are capped at 65\% of 1990 levels for the first modelling horizon of 2025 (compared to recorded emissions at 63\% of 1990 levels in 2023 \cite{europeancommission-2024c}).
Thereafter the cap is reduced to 45\%, 10\% and 0\% of 1990 levels at the 2030, 2040 and 2050 planning horizons in accordance with committed or planned EU policy.
Caps are linearly interpolated at the 2035 and 2045 planning horizons.

\subsubsection*{Spatial and temporal resolution}

The electricity, heat, hydrogen and gas energy carriers are spatially resolved; oil, biomass, ammonia, methanol and \ch{CO2} are not.
The model is configured with a spatial scale of 50 nodes (see \cref{fig:methodology}) to represent the modelling region, which consists of all EU countries except Cyprus and Malta, while also including Albania, Montenegro, North Macedonia, Norway, Serbia, Switzerland and the United Kingdom; $k$-means clustering is used to cluster smaller regions corresponding to transmission substations down to 50 regions based renewable energy capacity factor and electricity load time series.
In the temporal dimension, we aggregated hourly time-steps (8760 over one year) down to 750 time-steps using a segmentation approach \cite{pineda-morales-2018}, which is known to be more accurate than the equivalent $\sim$11.6-hourly uniform aggregation \cite{kotzur-markewitz-ea-2018a}.

Results have been validated against higher-resolution models under two different scenarios (level (a) and (c) for CCS potential, with the other settings being medium biomass availability, limited green fuel imports, baseline electrolyser costs and baseline transportation electrification) and with a 5\% slack level.
Across the two scenarios, cost optimisations as well as green hydrogen min- and maximisation and over the 2040, 2045 and 2050 planning horizons, we observe a maximum relative error of 3.7\% across all evaluated metrics for the selected 750 time-step 50 cluster resolution, as compared to a 2000 time-step 60 cluster model.
See \cref{supfig:resolution-errors} for the complete validation results.

\subsubsection*{Near-optimal modelling}

In order to explore different options for green hydrogen production, we exploit near-optimal solutions.
Near-optimal modelling was first applied to energy systems modelling by DeCarolis \cite{decarolis-2011} in 2011 under the name \emph{Modelling to Generate Alternatives} (MGA) and has subsequently been applied in various contexts \cite{neumann-brown-2021,neumann-brown-2023,grochowicz-vangreevenbroek-ea-2023,pedersen-victoria-ea-2021,vangreevenbroek-grochowicz-ea-2023} to reveal the range of options that are available for energy system designs when cost-optimality is relaxed slightly.

We develop a novel methodology to apply the near-optimal perspective to sequential optimisation over multiple planning horizons (``myopic foresight multi-horizon optimisation'').
This methodology is well-suited to studies where individual technologies are min- and maximised in order to explore the extremes of the near-optimal space in certain dimensions; it does not immediately generalise to other approaches such as the hop-skip-jump method \cite{decarolis-2011} or full explorations of the near-optimal space in multiple dimensions \cite{grochowicz-vangreevenbroek-ea-2023}.

Following Grochowicz et al.~\cite{grochowicz-vangreevenbroek-ea-2023}, let $\mathcal{F}_\varepsilon = \{x \in \mathbb{R}^n \mid Ax \leq b \text{ and } c \cdot x \leq (1 + \varepsilon) \cdot c^*\}$ be the $\varepsilon$-near-optimal space of an energy system model defined by the linear program $\min c \cdot x \text{ s.t. } Ax \leq b$ with optimal value $c^*$.
For multiple planning horizons with capacities carried over from one to the next, let $x_1^*, x_2^*, \dots$ be cost-optimal solutions at planning horizons $1, 2, \dots$, and let $x_0^*$ be the state of the current energy infrastructure before the first planning horizon.
Then the capacity expansion problem at horizon $i$ depends on $x_{i-1}^*$; let $\min c_i x_i \text{ s.t. } A_{i \mid x_{i-1}^*} x_i \leq b_{i \mid x_{i-1}^*}$ be the corresponding linear program.
We define the $\varepsilon$-near-optimal space at the $i$th planning horizon as
\begin{equation}
  \begin{split}
    \mathcal{F}_\varepsilon^{(i)} = \left\{x_i \in \mathbb{R}^n \mid \exists x_{i-1} \in \mathcal{F}_\varepsilon^{(i-1)} \text{ s.t. } A_{i \mid x_{i-1}} x_i \leq b_{i \mid x_{i-1}} \right.\\
    \left. \text{ and } c_i \cdot x_i \leq (1 + \varepsilon) \cdot c_i^* \vphantom{\mathcal{F}^{(i)}_\varepsilon}\right\}.
  \end{split}
\end{equation}
That is, each element in $\mathcal{F}_\varepsilon^{(i)}$ must be near-optimal with respect to optimal sequence of solutions $x_1^*, x_2^*, \dots$ (hence the constraint $c_i \cdot x_i \leq (1 + \varepsilon) \cdot c_i^*$), and must be feasible in the linear program $A_{i \mid x_{i-1}} x_i \leq b_{i \mid x_{i-1}}$ based on \emph{some} solution $x_{i-1}$ in the near-optimal space $\mathcal{F}_\varepsilon^{(i-1)}$ from previous horizon.

Instead of working with the full-dimensional spaces $\mathcal{F}_\varepsilon^{(i)}$, we project down to the single variable of interest: the annual sum of green hydrogen production (specifically, the sum of hydrogen production from electrolysis) across the entire model.
Under this projection, the image of $\mathcal{F}_\varepsilon^{(i)}$ is a line segment.

Mapping out $\mathcal{F}_\varepsilon^{(i)}$ is difficult because there is no obvious way of finding out which points in $\mathcal{F}_\varepsilon^{(i-1)}$ lead to the extremes of $\mathcal{F}_\varepsilon^{(i)}$.
In this study, we find points contained in $\mathcal{F}_\varepsilon^{(i)}$ that are obtained by max- and minimising green hydrogen production at each time horizon.
Let $h \in \mathbb{R}^n$ be the vector such that $h \cdot x$ is the total amount of hydrogen produced by electrolysis in $x$.
Then we define $x_0^{\text{max}} = x_0$ and for $i \geq 1$:
\begin{equation*}
  x_i^{\text{max}} = \argmax_{x \in \mathbb{R}^n} h \cdot x \text{ s.t. } A_{i \mid x_{i-1}^{\text{max}}} x \leq b_{i \mid x_{i-1}^{\text{max}}} \text{ and } c_i \cdot x \leq (1 + \varepsilon) \cdot c_i^*.
\end{equation*}
We define $x_i^{\text{min}}$ analogously.
That is, solutions $x_i^{\text{max}}$ and $x_i^{\text{min}}$ maximising and minimising green hydrogen production at time horizon $i$, respectively, are based on the corresponding solutions at horizon $i-1$ but must be near-optimal with respect to the cost-optimal sequence $x_1, x_2, \dots$.
In particular, capacities that do not reach the end of their lifetime are carrier over from $x_{i-1}^{\text{max}}$ to $x_i^{\text{max}}$ (and likely for minimisations).
The main results in this study are based on the sequences $x_1^{\text{max}}, x_2^{\text{max}}, \dots$ and $x_1^{\text{min}}, x_2^{\text{min}}, \dots$ in a number of different scenarios and for three different values of $\varepsilon$.

At each planning horizon, $x_i^{\text{max}}$ is not guaranteed to be the overall solution with maximum green hydrogen production in $\mathcal{F}_\varepsilon^{(i)}$, since this overall maximum may not be based on $x_{i-1}^{\text{max}}$.
As such, the near-optimal ranges of green hydrogen production we find in the present study represent conservative estimates, whereas the theoretical maximum ranges could be wider still.

Alternative extensions of near-optimal methods to a myopic multi-horizon planning regime are possible.
One interesting variation on the above definition would be one where total system cost slack could be distributed non-uniformly over time-horizons.
In particular, one could consider definitions with a total cost slack $\varepsilon \cdot \sum_i c_i^*$ summed over all planning horizons, distributed in various ways over near-optimal solutions at different planning horizons.
This could represent, for instance, the possibility to save in earlier time horizons only to spend the savings in later time horizons, or to spend borrowed funds early and cut spending later.
These dynamics are not representable in our study --- it is possible that greater spending flexibility leads to wider near-optimal spaces, making our proposed green hydrogen target near-optimal for more scenarios.
However, the exact effects are difficult to predict.

\subsubsection*{Robust corridors}

\Cref{fig:robustness} shows robust corridors of green hydrogen production.
At each time horizon, the width of the robust corridor is calculated as the intersection of the projected near-optimal spaces of all scenarios under consideration.
Indeed, let $S$ be a set of scenarios, and $I_{i, s} = [x_{i, s}^{\text{min}}, x_{i, s}^{\text{max}}]$ be the near-optimal range of green hydrogen production in scenario $s \in S$ and planning horizon $i$.
Then we define the \emph{robust} range of green hydrogen production at planning horizon $i$ to be the intersection $R_i = \bigcap_{s \in S} I_{i, s}$. This is what is shown in the darkest shade of green in \cref{fig:robustness} at successive time horizons.
Of course, $R_i$ may be empty; this is more likely at lower $\varepsilon$.

The quantiles that are also shown in \cref{fig:robustness} indicate ranges of points that are contained in a certain fraction of intervals $I_{i, s}$ for $s \in S$. Specifically, for $0 \leq q \leq 1$, we can define $R_{i,q} = \left\{x \in \mathbb{R} \text{ s.t. } |\{s \in S \mid x \in I_{i, s}\}| / |S| \geq q\right\}$ such that $R_i = R_{i,1}$ and, for example, $R_{i, 0.75}$ contains those points that are contained in at least 75\% of the intervals $I_{i, s}$ for $s \in S$.
\Cref{fig:robustness} shows $R_{i, 0.9}$ and $R_{i, 0.75}$ in addition to $R_i$ for different sets of scenarios $S$ and over planning horizons $i=2030, \dots, 2050$.

\subsubsection*{Green hydrogen subsidy calculations}

For scenarios where cost-optimal production of green hydrogen in 2040 is below the target of \qty{25}{Mt}, we would like to calculate the subsidy level required to reach the target.
This is the price that would have to be paid by a central authority in order for the target to be reached; total subsidy volume does not necessarily correspond directly to the total system cost slack. 

For a given scenario in which cost-optimal green hydrogen production is less than \qty{25}{Mt}, it can be shown that there exists a subsidy level $t$ (in \qty{}{EUR/kg\ch{H2}} of produced green hydrogen) such that producing \qty{25}{Mt} of green hydrogen is cost-optimal given the subsidy $t$ \cite{finke-weber-ea-2024}.
The total subsidy volume in EUR is then obtained by multiplying $t$ by \qty{25}{Mt}.

In fact, suppose we generate a near-optimal solution with some cost slack $\varepsilon$ by maximising green hydrogen production, and find that $M$ tonnes of green hydrogen are produced in this near-optimal solution.
Then, letting $\mu$ be the dual variable to the total system cost constraint in the near-optimal model, the subsidy level (in EUR/kg\ch{H2}) required to make the production of $M$ tonnes of green hydrogen cost-optimal is equal (up to a change in units from MWh to kg\ch{H2} using the lower heating value) to $1/\mu$ (assuming the total system cost constraint is binding so that $\mu \neq 0$) \cite{finke-weber-ea-2024}.

In our case, we have results for three different slack levels ($\varepsilon = 2\%, 5\%, 10\%$) resulting in different amounts of maximised green hydrogen production; we additionally count cost-optimisations as having a slack of $0\%$.
In order to estimate the subsidy level required to induce the \qty{25}{Mt} target, suppose $\varepsilon_1, \varepsilon_2$ are two consecutive slack levels (including $0\%$ as an option) with green hydrogen production amounts of $M_1$ and $M_2$ (in Mt) such that $M_1 < 25 < M_2$.
Let $\mu_1, \mu_2$ be the dual variables of the respective total system cost constraints, taken to be $\inf$ for $\varepsilon = 0\%$.
Then we estimate the subsidy $t$ required to induce \qty{25}{Mt} of green hydrogen to be
\begin{equation*}
  t \approx \frac{M_2 - 25}{M_2 - M_1} \cdot \frac{1}{\mu_1} + \frac{25 - M_1}{M_2 - M_1} \cdot \frac{1}{\mu_2},
\end{equation*}
which is a linear interpolation between $1/\mu_1$ and $1/\mu_2$ based on where the \qty{25}{Mt} target is relative to $M_1$ and $M_2$.
If \qty{25}{Mt} is greater than the maximum green hydrogen production at a $10\%$ cost slack, we fall back on $1/\mu$ at $\varepsilon=10\%$ as our estimated subsidy level.
The average figure of \qty{0.55}{EUR/kg\ch{H2}} given in the main text is the average of $t$ over all scenarios, taking $t$ to be 0 in scenarios where the cost-optimal amount of green hydrogen is more than \qty{25}{Mt}.

\subsubsection*{Sensitivity analysis}

\Cref{tab:sensitivity} records the sensitivity of green hydrogen production to the five main scenario settings as well as three CCS-specific parameters.
Recall that we consider five different categories of scenario settings:
\begin{enumerate}
  \item CCS potential,
  \item Biomass potential,
  \item Green fuel imports,
  \item Electrolyser capital cost,
  \item Weather year.
\end{enumerate}
Each category has two or three settings, labelled \textbf{(a)}, \textbf{(b)} and, where applicable, \textbf{(c)}.

In order to compute sensitivity coefficients, we map the categories to dummy variables $x_1, x_2, x_3, x_4, x_5, x_6$.
Then, the coefficients are given as the coefficients of the best-fit linear function
\begin{equation}
    y = a_1 x_1 + a_2 x_2 + a_3 x_3 + a_4 x_4 + a_5 x_5 + a_6 x_6
\end{equation}
where $y$ is total green hydrogen production.
In order to perform a linear regression, we map the above settings to numerical values of the dummy variables using the map $\{(a) \mapsto 0, (b) \mapsto 1\}$ or $\{(a) \mapsto 0, (b) \mapsto 0.5, (c) \mapsto 1\}$ depending on whether the category has two or three different settings.
Each of the $216 = 3 \cdot 3 \cdot 3 \cdot 2 \cdot 2 \cdot 2$ different combinations of scenario settings, using the above mapping together with modelled total green hydrogen production, thus produces a point in $\mathbb{R^5 \times \mathbb{R}}$.
We perform the sensitivity analysis over the 2040, 2045 and 2050 planning horizons, making for a total of 648 points.
A linear regression on these points gives the desired coefficients $a_1, \dots, a_6$.
This whole analysis is repeated three times: once for each of green hydrogen minimisations (at 5\% slack), cost-optimisations and green hydrogen maximisations (at 5\% slack), corresponding to the three columns in \cref{tab:sensitivity}.

A separate set of optimisations is run in order to investigate the individual effects of the three different parameters making up the ``CCS potential'' category of settings.
The model used for this investigation is identical to that used in the main analysis, but the set of scenarios consists of all 27 combinations of the following settings:
\begin{enumerate}
\item \ch{CO2} sequestration potential:\\
  \textbf{(a)} 25/125/\qty{275}{Mt/a} by 2030/2040/2050\\
  \textbf{(b)} 50/250/\qty{550}{Mt/a} by 2030/2040/2050\\
  \textbf{(c)} 100/500/\qty{1100}{Mt/a} by 2030/2040/2050
\item Cost of \ch{CO2} sequestration:\\
  \textbf{(a)} \qty{30}{EUR/t\ch{CO2}} \\
  \textbf{(b)} \qty{20}{EUR/t\ch{CO2}} \\
  \textbf{(c)} \qty{15}{EUR/t\ch{CO2}}
\item Capital cost of carbon capture: \\
  \textbf{(a)} $+50\%$ \\
  \textbf{(b)} $+0\%$ \\
  \textbf{(c)} $-50\%$
\end{enumerate}
Apart from that, the weather year for this model is fixed to 1987, electrolyser cost is set to the baseline assumption and imports of green fuels are restricted not to exceed the volume of European green hydrogen production (corresponding to setting \textbf{(a)} for imports in the main set of scenarios).
The same procedure as above is used to compute the sensitivity coefficients shown in \cref{tab:sensitivity}.

\subsubsection*{Build year aggregation}

PyPSA-Eur was modified for the purposes of this study by aggregating components by build-year for the purposes of saving on computational time and memory requirements.
At the time of writing, PyPSA-Eur implements multi-horizon optimisations with myopic foresight by successively adding new, expandable components at each planning horizon and re-solving the model; components built in previous planning horizons are left intact but are not expandable.
(Components at the end of their lifetime are phased out.)
This leads to an accumulation of components in later planning horizons, increasing the time and especially memory footprint of optimisations.

However, groups of components that are built at different planning horizons but identical otherwise (for example, all wind power generators at a specific location) typically will be operated identically too.
Therefore, such groups of components can be aggregated before optimising the model without changing the results.
We implement this aggregation in PyPSA-Eur\footnote{PR has been submitted to the upstream codebase: \url{https://github.com/PyPSA/pypsa-eur/pull/1056}}; by aggregating just before each optimisation and disaggregating after each optimisation, other logic related to myopic planning (phasing out old components, etc.) is not affected.

Some components have properties that change over time; the primary example being improving efficiencies over time.
These components cannot be aggregated without some loss of accuracy in the results.
In this study, we exclude electrolysers and process emission carbon capture components from the aggregation procedure; both have improving efficiencies over time.
We validate the accuracy of optimisation results by running two sets of optimisations resembling those used for the main results but at lower (100 time-step) temporal resolution: one with and one without build year aggregation.
We find that build year aggregation leads to an average error of only 0.04\% in total green hydrogen production, and a maximum average error of 0.34\% across other model statistics including total system cost, total installed renewables, captured \ch{CO2} and electrolysis capacity factor.
As such, the aggregation leads to more than enough accuracy to support all our conclusions.
Memory consumption of optimisations at later planning horizons was improved by more than a factor of 2.

\bibliographystyle{unsrtnat}

\begin{thebibliography}{65}
\providecommand{\natexlab}[1]{#1}
\providecommand{\url}[1]{\texttt{#1}}
\expandafter\ifx\csname urlstyle\endcsname\relax
  \providecommand{\doi}[1]{doi: #1}\else
  \providecommand{\doi}{doi: \begingroup \urlstyle{rm}\Url}\fi

\bibitem[{European Commission}(2022)]{europeancommission-2022}
{European Commission}.
\newblock Commission staff working document implementing the {{REPowerEU}}
  action plan: Investment needs, hydrogen accelerator and achieving the
  bio-methane targets [{{SWD}}/2022/230 final], May 2022.
\newblock URL
  \url{https://eur-lex.europa.eu/legal-content/EN/TXT/?uri=SWD%3A2022%3A230%3AFIN}.

\bibitem[{European Commission}(2024{\natexlab{a}})]{europeancommission-2024b}
{European Commission}.
\newblock Securing our future {{Europe}}'s 2040 climate target and path to
  climate neutrality by 2050 building a sustainable, just and prosperous
  society, 2024{\natexlab{a}}.
\newblock URL
  \url{https://eur-lex.europa.eu/legal-content/EN/TXT/?uri=COM%3A2024%3A63%3AFIN}.

\bibitem[{The European Parliament and the Council of the European
  Union}(2021)]{eu-climate-law-2021}
{The European Parliament and the Council of the European Union}.
\newblock Regulation ({{EU}}) 2021/1119 of the {{European Parliament}} and of
  the {{Council}} of 30~{{June}} 2021 establishing the framework for achieving
  climate neutrality and amending {{Regulations}} ({{EC}}) {{No}}~401/2009 and
  ({{EU}}) 2018/1999 (`{{European Climate Law}}'), June 2021.
\newblock URL \url{http://data.europa.eu/eli/reg/2021/1119/oj/eng}.

\bibitem[B{\'e}res et~al.(2024)B{\'e}res, Nijs, Boldrini, and {van den
  Broek}]{beres-nijs-ea-2024}
Rebeka B{\'e}res, Wouter Nijs, Annika Boldrini, and Machteld {van den Broek}.
\newblock Will hydrogen and synthetic fuels energize our future? {{Their}} role
  in {{Europe}}'s climate-neutral energy system and power system dynamics.
\newblock \emph{Applied Energy}, 375:\penalty0 124053, December 2024.
\newblock ISSN 0306-2619.
\newblock \doi{10.1016/j.apenergy.2024.124053}.

\bibitem[Seck et~al.(2022)Seck, Hache, Sabathier, Guedes, Reigstad, Straus,
  Wolfgang, Ouassou, Askeland, Hjorth, Skjelbred, Andersson, Douguet,
  Villavicencio, Tr{\"u}by, Brauer, and Cabot]{seck-hache-ea-2022}
Gondia~S. Seck, Emmanuel Hache, Jerome Sabathier, Fernanda Guedes, Gunhild~A.
  Reigstad, Julian Straus, Ove Wolfgang, Jabir~A. Ouassou, Magnus Askeland, Ida
  Hjorth, Hans~I. Skjelbred, Leif~E. Andersson, Sebastien Douguet, Manuel
  Villavicencio, Johannes Tr{\"u}by, Johannes Brauer, and Clement Cabot.
\newblock Hydrogen and the decarbonization of the energy system in europe in
  2050: {{A}} detailed model-based analysis.
\newblock \emph{Renewable and Sustainable Energy Reviews}, 167:\penalty0
  112779, October 2022.
\newblock ISSN 13640321.
\newblock \doi{10.1016/j.rser.2022.112779}.

\bibitem[Neumann et~al.(2023)Neumann, Zeyen, Victoria, and
  Brown]{neumann-zeyen-ea-2023}
Fabian Neumann, Elisabeth Zeyen, Marta Victoria, and Tom Brown.
\newblock The potential role of a hydrogen network in {{Europe}}.
\newblock \emph{Joule}, 7\penalty0 (8):\penalty0 1793--1817, August 2023.
\newblock ISSN 2542-4351.
\newblock \doi{10.1016/j.joule.2023.06.016}.

\bibitem[Mignone et~al.(2024)Mignone, Clarke, Edmonds, Gurgel, Herzog, Johnson,
  Mallapragada, McJeon, Morris, O'Rourke, Paltsev, Rose, Steinberg, and
  Venkatesh]{mignone-clarke-ea-2024}
Bryan~K. Mignone, Leon Clarke, James~A. Edmonds, Angelo Gurgel, Howard~J.
  Herzog, Jeremiah~X. Johnson, Dharik~S. Mallapragada, Haewon McJeon, Jennifer
  Morris, Patrick~R. O'Rourke, Sergey Paltsev, Steven~K. Rose, Daniel~C.
  Steinberg, and Aranya Venkatesh.
\newblock Drivers and implications of alternative routes to fuels
  decarbonization in net-zero energy systems.
\newblock \emph{Nature Communications}, 15\penalty0 (1):\penalty0 3938, May
  2024.
\newblock ISSN 2041-1723.
\newblock \doi{10.1038/s41467-024-47059-0}.

\bibitem[Neumann et~al.(2024)Neumann, Hampp, and Brown]{neumann-hampp-ea-2024}
Fabian Neumann, Johannes Hampp, and Tom Brown.
\newblock Energy {{Imports}} and {{Infrastructure}} in a {{Carbon-Neutral
  European Energy System}}, April 2024.

\bibitem[Ganter et~al.(2024)Ganter, Gabrielli, and
  Sansavini]{ganter-gabrielli-ea-2024}
Alissa Ganter, Paolo Gabrielli, and Giovanni Sansavini.
\newblock Near-term infrastructure rollout and investment strategies for
  net-zero hydrogen supply chains.
\newblock \emph{Renewable and Sustainable Energy Reviews}, 194:\penalty0
  114314, April 2024.
\newblock ISSN 13640321.
\newblock \doi{10.1016/j.rser.2024.114314}.

\bibitem[Blanco et~al.(2018)Blanco, Nijs, Ruf, and Faaij]{blanco-nijs-ea-2018}
Herib Blanco, Wouter Nijs, Johannes Ruf, and Andr{\'e} Faaij.
\newblock Potential for hydrogen and {{Power-to-Liquid}} in a low-carbon {{EU}}
  energy system using cost optimization.
\newblock \emph{Applied Energy}, 232:\penalty0 617--639, December 2018.
\newblock ISSN 0306-2619.
\newblock \doi{10.1016/j.apenergy.2018.09.216}.

\bibitem[Wu et~al.(2022)Wu, M{\"u}ller, and Pfenninger]{wu-muller-ea-2022}
Fei Wu, Adrian M{\"u}ller, and Stefan Pfenninger.
\newblock Strategic uses for ancillary bioenergy in a carbon-neutral and
  fossil-free 2050 {{European}} energy system.
\newblock \emph{Environmental Research Letters}, 2022.
\newblock ISSN 1748-9326.
\newblock \doi{10.1088/1748-9326/aca9e1}.

\bibitem[Millinger et~al.(2023)Millinger, Hedenus, Reichenberg, Zeyen, Neumann,
  and Berndes]{millinger-hedenus-ea-2023}
Markus Millinger, Fredrik Hedenus, Lina Reichenberg, Elisabeth Zeyen, Fabian
  Neumann, and G{\"o}ran Berndes.
\newblock Diversity of biomass usage pathways to achieve emissions targets in
  the {{European}} energy system, July 2023.
\newblock ISSN 2693-5015.

\bibitem[Pickering et~al.(2022)Pickering, Lombardi, and
  Pfenninger]{pickering-lombardi-ea-2022}
Bryn Pickering, Francesco Lombardi, and Stefan Pfenninger.
\newblock Diversity of options to eliminate fossil fuels and reach carbon
  neutrality across the entire {{European}} energy system.
\newblock \emph{Joule}, 6\penalty0 (6):\penalty0 1253--1276, June 2022.
\newblock ISSN 25424351.
\newblock \doi{10.1016/j.joule.2022.05.009}.

\bibitem[Zeyen et~al.(2023)Zeyen, Victoria, and Brown]{zeyen-victoria-ea-2023}
Elisabeth Zeyen, Marta Victoria, and Tom Brown.
\newblock Endogenous learning for green hydrogen in a sector-coupled energy
  model for {{Europe}}.
\newblock \emph{Nature Communications}, 14\penalty0 (1):\penalty0 3743, June
  2023.
\newblock ISSN 2041-1723.
\newblock \doi{10.1038/s41467-023-39397-2}.

\bibitem[Fleiter et~al.(2024)Fleiter, Fragoso, Lux, Aliba{\c s}, {Al-Dabbas},
  Manz, Neuner, Wei{\ss}enburger, Rehfeldt, and
  Sensfu{\ss}]{fleiter-fragoso-ea-2024}
Tobias Fleiter, Joshua Fragoso, Benjamin Lux, {\c S}irin Aliba{\c s}, Khaled
  {Al-Dabbas}, Pia Manz, Felix Neuner, Bastian Wei{\ss}enburger, Matthias
  Rehfeldt, and Frank Sensfu{\ss}.
\newblock Hydrogen infrastructure in the future {{CO2-neutral}} european energy
  system---how does the demand for hydrogen affect the need for infrastructure?
\newblock \emph{Energy Technology}, 2024.
\newblock \doi{10.1002/ente.202300981}.

\bibitem[Ruiz et~al.(2019)Ruiz, Nijs, Tarvydas, Sgobbi, Zucker, Pilli, Camia,
  Thiel, {Hoyer-Klick}, Dalla, Kober, Badger, Volker, Elbersen, Brosowski,
  Thr{\"a}n, and Jonsson]{ruiz-nijs-ea-2019}
Castello~Pablo Ruiz, Wouter Nijs, Dalius Tarvydas, Alessandra Sgobbi, Andreas
  Zucker, Roberto Pilli, Andrea Camia, Christian Thiel, Carsten {Hoyer-Klick},
  Longa~Francesco Dalla, Tom Kober, Jake Badger, Patrick Volker, Berien
  Elbersen, Andre Brosowski, Daniela Thr{\"a}n, and Klas Jonsson.
\newblock {{ENSPRESO}} - an open data, {{EU-28}} wide, transparent and coherent
  database of wind, solar and biomass energy potentials, June 2019.
\newblock URL
  \url{https://publications.jrc.ec.europa.eu/repository/handle/JRC116900}.

\bibitem[{European Commission} and {Directorate-General for
  Energy}(2024)]{europeancommission-directorate-generalforenergy-2024}
{European Commission} and {Directorate-General for Energy}.
\newblock Union bioenergy sustainability report -- {{Study}} to support
  reporting under {{Article}} 35 of {{Regulation}} ({{EU}}) 2018/1999 --
  {{Final}} report.
\newblock Technical report, Publications Office of the European Union, 2024.

\bibitem[Galimova et~al.(2023)Galimova, Ram, Bogdanov, Fasihi, Gulagi, Khalili,
  and Breyer]{galimova-ram-ea-2023}
Tansu Galimova, Manish Ram, Dmitrii Bogdanov, Mahdi Fasihi, Ashish Gulagi,
  Siavash Khalili, and Christian Breyer.
\newblock Global trading of renewable electricity-based fuels and chemicals to
  enhance the energy transition across all sectors towards sustainability.
\newblock \emph{Renewable and Sustainable Energy Reviews}, 183:\penalty0
  113420, September 2023.
\newblock ISSN 1364-0321.
\newblock \doi{10.1016/j.rser.2023.113420}.

\bibitem[M{\"u}ller et~al.(2024)M{\"u}ller, Riemer, and
  Eckstein]{muller-riemer-ea-2024}
Viktor~Paul M{\"u}ller, Matia Riemer, and Johannes Eckstein.
\newblock Are {{Carbon-Based E-Fuels}} a {{Viable Path}} to
  {{Decarbonization}}? - {{A Critical Comparison}} of {{Global Supply}} and
  {{Demand}}.
\newblock In \emph{2024 20th {{International Conference}} on the {{European
  Energy Market}} ({{EEM}})}, pages 1--7, June 2024.
\newblock \doi{10.1109/EEM60825.2024.10608862}.

\bibitem[Odenweller et~al.(2022)Odenweller, Ueckerdt, Nemet, Jensterle, and
  Luderer]{odenweller-ueckerdt-ea-2022}
Adrian Odenweller, Falko Ueckerdt, Gregory~F. Nemet, Miha Jensterle, and Gunnar
  Luderer.
\newblock Probabilistic feasibility space of scaling up green hydrogen supply.
\newblock \emph{Nature Energy}, 7\penalty0 (9):\penalty0 854--865, September
  2022.
\newblock ISSN 2058-7546.
\newblock \doi{10.1038/s41560-022-01097-4}.

\bibitem[Odenweller and Ueckerdt(2025)]{odenweller-ueckerdt-2025}
Adrian Odenweller and Falko Ueckerdt.
\newblock The green hydrogen ambition and implementation gap.
\newblock \emph{Nature Energy}, 10\penalty0 (1):\penalty0 110--123, January
  2025.
\newblock ISSN 2058-7546.
\newblock \doi{10.1038/s41560-024-01684-7}.

\bibitem[{European Commission}(2024{\natexlab{b}})]{europeancommission-2024}
{European Commission}.
\newblock Impact {{Assessment}} of the 2040 {{Climate Target}}.
\newblock Technical Report SWD/2024/63, February 2024{\natexlab{b}}.

\bibitem[{Jacopo Maria Pepe} et~al.(2023){Jacopo Maria Pepe}, {Dawud Ansari},
  and {Rosa Melissa Gehrung}]{jacopomariapepe-dawudansari-ea-2023}
{Jacopo Maria Pepe}, {Dawud Ansari}, and {Rosa Melissa Gehrung}.
\newblock {The Geopolitics of Hydrogen}.
\newblock Technical report, Stiftung Wissenschaft und Politik, 2023.
\newblock URL
  \url{https://www.swp-berlin.org/publikation/the-geopolitics-of-hydrogen}.

\bibitem[{The European Parliament and the Council of the European
  Union}(2024)]{eu-net-zero-industry-2024}
{The European Parliament and the Council of the European Union}.
\newblock The {{Net-Zero Industry Act}}, June 2024.
\newblock URL \url{http://data.europa.eu/eli/reg/2024/1735/oj}.

\bibitem[Kazlou et~al.(2024)Kazlou, Cherp, and Jewell]{kazlou-cherp-ea-2024}
Tsimafei Kazlou, Aleh Cherp, and Jessica Jewell.
\newblock Feasible deployment of carbon capture and storage and the
  requirements of climate targets.
\newblock \emph{Nature Climate Change}, September 2024.
\newblock ISSN 1758-6798.
\newblock \doi{10.1038/s41558-024-02104-0}.

\bibitem[Anthonsen and Christensen(2021)]{anthonsen-christensen-2021}
Karen~Lyng Anthonsen and Niels~Peter Christensen.
\newblock {{EU Geological CO}}{$_2$} storage summary.
\newblock Technical Report 2021/34, {Geological Survey of Denmark and
  Greenland}, 2021.

\bibitem[Quarton et~al.(2020)Quarton, Tlili, Welder, Mansilla, Blanco,
  Heinrichs, Leaver, Samsatli, Lucchese, Robinius, and
  Samsatli]{quarton-tlili-ea-2020}
Christopher~J. Quarton, Olfa Tlili, Lara Welder, Christine Mansilla, Herib
  Blanco, Heidi Heinrichs, Jonathan Leaver, Nouri~J. Samsatli, Paul Lucchese,
  Martin Robinius, and Sheila Samsatli.
\newblock The curious case of the conflicting roles of hydrogen in global
  energy scenarios.
\newblock \emph{Sustainable Energy \& Fuels}, 4\penalty0 (1):\penalty0 80--95,
  2020.
\newblock ISSN 2398-4902.
\newblock \doi{10.1039/C9SE00833K}.

\bibitem[Schreyer et~al.(2024)Schreyer, Ueckerdt, Pietzcker, Rodrigues,
  Rottoli, Madeddu, Pehl, Hasse, and Luderer]{schreyer-ueckerdt-ea-2024}
Felix Schreyer, Falko Ueckerdt, Robert Pietzcker, Renato Rodrigues, Marianna
  Rottoli, Silvia Madeddu, Michaja Pehl, Robin Hasse, and Gunnar Luderer.
\newblock Distinct roles of direct and indirect electrification in pathways to
  a renewables-dominated {{European}} energy system.
\newblock \emph{One Earth}, 7\penalty0 (2):\penalty0 226--241, February 2024.
\newblock ISSN 2590-3330, 2590-3322.
\newblock \doi{10.1016/j.oneear.2024.01.015}.

\bibitem[Kountouris et~al.(2024)Kountouris, Bramstoft, Madsen,
  {Gea-Berm{\'u}dez}, M{\"u}nster, and Keles]{kountouris-bramstoft-ea-2024}
Ioannis Kountouris, Rasmus Bramstoft, Theis Madsen, Juan {Gea-Berm{\'u}dez},
  Marie M{\"u}nster, and Dogan Keles.
\newblock A unified {{European}} hydrogen infrastructure planning to support
  the rapid scale-up of hydrogen production.
\newblock \emph{Nature Communications}, 15\penalty0 (1):\penalty0 5517, June
  2024.
\newblock ISSN 2041-1723.
\newblock \doi{10.1038/s41467-024-49867-w}.

\bibitem[Victoria et~al.(2022)Victoria, Zeyen, and
  Brown]{victoria-zeyen-ea-2022}
Marta Victoria, Elisabeth Zeyen, and Tom Brown.
\newblock Speed of technological transformations required in {{Europe}} to
  achieve different climate goals.
\newblock \emph{Joule}, 6\penalty0 (5):\penalty0 1066--1086, May 2022.
\newblock ISSN 2542-4785, 2542-4351.
\newblock \doi{10.1016/j.joule.2022.04.016}.

\bibitem[{European Environment Agency}(2023)]{europeanenvironmentagency-2023a}
{European Environment Agency}.
\newblock Scientific advice for the determination of an {{EU-wide}} 2040
  climate target and a greenhouse gas budget for 2030--2050.
\newblock Technical report, Publications Office of the European Union, June
  2023.
\newblock URL \url{https://data.europa.eu/doi/10.2800/609405}.

\bibitem[Sgouridis et~al.(2022)Sgouridis, Kimmich, Sol{\'e}, {\v C}ern{\'y},
  Ehlers, and Kerschner]{sgouridis-kimmich-ea-2022a}
Sgouris Sgouridis, Christian Kimmich, Jordi Sol{\'e}, Martin {\v C}ern{\'y},
  Melf-Hinrich Ehlers, and Christian Kerschner.
\newblock Visions before models: {{The}} ethos of energy modeling in an era of
  transition.
\newblock \emph{Energy Research \& Social Science}, 88:\penalty0 102497, June
  2022.
\newblock ISSN 2214-6296.
\newblock \doi{10.1016/j.erss.2022.102497}.

\bibitem[Trutnevyte(2016)]{trutnevyte-2016}
Evelina Trutnevyte.
\newblock Does cost optimization approximate the real-world energy transition?
\newblock \emph{Energy}, 106:\penalty0 182--193, July 2016.
\newblock ISSN 0360-5442.
\newblock \doi{10.1016/j.energy.2016.03.038}.

\bibitem[DeCarolis(2011)]{decarolis-2011}
Joseph~F. DeCarolis.
\newblock Using modeling to generate alternatives ({{MGA}}) to expand our
  thinking on energy futures.
\newblock \emph{Energy Economics}, 33\penalty0 (2):\penalty0 145--152, March
  2011.
\newblock ISSN 01409883.
\newblock \doi{10.1016/j.eneco.2010.05.002}.

\bibitem[{van Greevenbroek} et~al.(2023){van Greevenbroek}, Grochowicz,
  Zeyringer, and Benth]{vangreevenbroek-grochowicz-ea-2023}
Koen {van Greevenbroek}, Aleksander Grochowicz, Marianne Zeyringer, and
  Fred~Espen Benth.
\newblock Enabling agency: Trade-offs between regional and integrated energy
  systems design flexibility, December 2023.

\bibitem[Neumann and Brown(2021)]{neumann-brown-2021}
Fabian Neumann and Tom Brown.
\newblock The near-optimal feasible space of a renewable power system model.
\newblock \emph{Electric Power Systems Research}, 190:\penalty0 106690, January
  2021.
\newblock ISSN 0378-7796.
\newblock \doi{10.1016/j.epsr.2020.106690}.

\bibitem[Neumann and Brown(2023)]{neumann-brown-2023}
Fabian Neumann and Tom Brown.
\newblock Broad {{Ranges}} of {{Investment Configurations}} for {{Renewable
  Power Systems}}, {{Robust}} to {{Cost Uncertainty}} and {{Near-Optimality}}.
\newblock \emph{iScience}, page 106702, April 2023.
\newblock ISSN 25890042.
\newblock \doi{10.1016/j.isci.2023.106702}.

\bibitem[Grochowicz et~al.(2023)Grochowicz, {van Greevenbroek}, Benth, and
  Zeyringer]{grochowicz-vangreevenbroek-ea-2023}
Aleksander Grochowicz, Koen {van Greevenbroek}, Fred~Espen Benth, and Marianne
  Zeyringer.
\newblock Intersecting near-optimal spaces: {{European}} power systems with
  more resilience to weather variability.
\newblock \emph{Energy Economics}, 118:\penalty0 106496, February 2023.
\newblock ISSN 0140-9883.
\newblock \doi{10.1016/j.eneco.2022.106496}.

\bibitem[Esser et~al.(2024)Esser, Finke, Bertsch, and
  L{\"o}schel]{esser-finke-ea-2024}
Katharina Esser, Jonas Finke, Valentin Bertsch, and Andreas L{\"o}schel.
\newblock Participatory {{Modelling}} to {{Generate Alternatives}} for
  {{Decarbonising Campus Energy Supply}}, December 2024.

\bibitem[Price and Keppo(2017)]{price-keppo-2017}
James Price and Ilkka Keppo.
\newblock Modelling to generate alternatives: {{A}} technique to explore
  uncertainty in energy-environment-economy models.
\newblock \emph{Applied Energy}, 195:\penalty0 356--369, June 2017.
\newblock ISSN 0306-2619.
\newblock \doi{10.1016/j.apenergy.2017.03.065}.

\bibitem[Sinha et~al.(2024)Sinha, Venkatesh, Jordan, Wade, Eshraghi, {de
  Queiroz}, Jaramillo, and Johnson]{sinha-venkatesh-ea-2024}
Aditya Sinha, Aranya Venkatesh, Katherine Jordan, Cameron Wade, Hadi Eshraghi,
  Anderson~R. {de Queiroz}, Paulina Jaramillo, and Jeremiah~X. Johnson.
\newblock Diverse decarbonization pathways under near cost-optimal futures.
\newblock \emph{Nature Communications}, 15\penalty0 (1):\penalty0 8165,
  September 2024.
\newblock ISSN 2041-1723.
\newblock \doi{10.1038/s41467-024-52433-z}.

\bibitem[H{\"o}rsch et~al.(2018)H{\"o}rsch, Hofmann, Schlachtberger, and
  Brown]{PyPSAEur}
Jonas H{\"o}rsch, Fabian Hofmann, David Schlachtberger, and Tom Brown.
\newblock {{PyPSA-Eur}}: {{An}} open optimisation model of the {{European}}
  transmission system.
\newblock \emph{Energy Strategy Reviews}, 22:\penalty0 207--215, November 2018.
\newblock ISSN 2211-467X.
\newblock \doi{10.1016/j.esr.2018.08.012}.

\bibitem[Brown et~al.(2018{\natexlab{a}})Brown, Schlachtberger, Kies, Schramm,
  and Greiner]{PyPSAEurSec}
T.~Brown, D.~Schlachtberger, A.~Kies, S.~Schramm, and M.~Greiner.
\newblock Synergies of sector coupling and transmission reinforcement in a
  cost-optimised, highly renewable {{European}} energy system.
\newblock \emph{Energy}, 160:\penalty0 720--739, October 2018{\natexlab{a}}.
\newblock ISSN 0360-5442.
\newblock \doi{10.1016/j.energy.2018.06.222}.

\bibitem[Hofmann et~al.(2011)Hofmann, Kafadar, and Wickham]{letter-value-plot}
Heike Hofmann, Karen Kafadar, and Hadley Wickham.
\newblock Letter-value plots: {{Boxplots}} for large data.
\newblock Technical report, had.co.nz, 2011.

\bibitem[Brown et~al.(2018{\natexlab{b}})Brown, {Bischof-Niemz}, Blok, Breyer,
  Lund, and Mathiesen]{brown-bischof-niemz-ea-2018}
T.W. Brown, T.~{Bischof-Niemz}, K.~Blok, C.~Breyer, H.~Lund, and B.V.
  Mathiesen.
\newblock Response to `{{Burden}} of proof: {{A}} comprehensive review of the
  feasibility of 100\% renewable-electricity systems'.
\newblock \emph{Renewable and Sustainable Energy Reviews}, 92:\penalty0
  834--847, September 2018{\natexlab{b}}.
\newblock ISSN 13640321.
\newblock \doi{10.1016/j.rser.2018.04.113}.

\bibitem[{European Court of Auditors}(2024)]{europeancourtofauditors-2024}
{European Court of Auditors}.
\newblock The {{EU}}'s industrial policy on renewable hydrogen -- {{Legal}}
  framework has been mostly adopted -- time for a reality check.
\newblock Technical Report 11/2024, 2024.
\newblock URL \url{https://www.eca.europa.eu/en/publications?ref=sr-2024-11}.

\bibitem[Zeyen et~al.(2025)Zeyen, Kalweit, Victoria, and
  Brown]{zeyen-kalweit-ea-2025}
Elisabeth Zeyen, Sina Kalweit, Marta Victoria, and Tom Brown.
\newblock Shifting burdens: How delayed decarbonisation of road transport
  affects other sectoral emission reductions.
\newblock \emph{Environmental Research Letters}, 20\penalty0 (4):\penalty0
  044044, March 2025.
\newblock ISSN 1748-9326.
\newblock \doi{10.1088/1748-9326/adc290}.

\bibitem[Commission(2022)]{europeancommission-2022a}
European Commission.
\newblock {{EU}} deal to end sale of new {{CO2}} emitting cars by 2035, 2022.
\newblock URL
  \url{https://ec.europa.eu/commission/presscorner/detail/en/ip_22_6462}.

\bibitem[Odenweller and Ueckerdt(2024)]{odenweller-ueckerdt-2024}
Adrian Odenweller and Falko Ueckerdt.
\newblock The green hydrogen ambition and implementation gap, June 2024.

\bibitem[Sand et~al.(2023)Sand, Skeie, Sandstad, Krishnan, Myhre, Bryant,
  Derwent, Hauglustaine, Paulot, Prather, and Stevenson]{sand-skeie-ea-2023}
Maria Sand, Ragnhild~Bieltvedt Skeie, Marit Sandstad, Srinath Krishnan, Gunnar
  Myhre, Hannah Bryant, Richard Derwent, Didier Hauglustaine, Fabien Paulot,
  Michael Prather, and David Stevenson.
\newblock A multi-model assessment of the {{Global Warming Potential}} of
  hydrogen.
\newblock \emph{Communications Earth \& Environment}, 4\penalty0 (1):\penalty0
  1--12, June 2023.
\newblock ISSN 2662-4435.
\newblock \doi{10.1038/s43247-023-00857-8}.

\bibitem[Hersbach et~al.(2018)Hersbach, Bell, Berrisford, Biavati, Hor{\'a}nyi,
  Mu{\~n}oz~Sabater, Nicolas, Peubey, Radu, Rozum, Schepers, Simmons, Soci,
  Dee, and Th{\'e}paut]{hersbach-bell-ea-2018}
Hans Hersbach, Bill Bell, Paul Berrisford, Gionata Biavati, And{\'a}s
  Hor{\'a}nyi, Joaqu{\'i}n Mu{\~n}oz~Sabater, Julien Nicolas, Carole Peubey,
  Raluca Radu, Iryna Rozum, Dinand Schepers, Adrian Simmons, Cornel Soci, Dick
  Dee, and Jean-No{\"e}l Th{\'e}paut.
\newblock {{ERA5}} hourly data on single levels from 1940 to present, 2018.

\bibitem[Smith et~al.(2021)Smith, Morris, Kheshgi, Teletzke, Herzog, and
  Paltsev]{smith-morris-ea-2021a}
Erin Smith, Jennifer Morris, Haroon Kheshgi, Gary Teletzke, Howard Herzog, and
  Sergey Paltsev.
\newblock The cost of {{CO2}} transport and storage in global integrated
  assessment modeling.
\newblock \emph{International Journal of Greenhouse Gas Control}, 109:\penalty0
  103367, July 2021.
\newblock ISSN 1750-5836.
\newblock \doi{10.1016/j.ijggc.2021.103367}.

\bibitem[Hampp et~al.(2023)Hampp, D{\"u}ren, and Brown]{hampp-duren-ea-2023}
Johannes Hampp, Michael D{\"u}ren, and Tom Brown.
\newblock Import options for chemical energy carriers from renewable sources to
  {{Germany}}.
\newblock \emph{PLOS ONE}, 18\penalty0 (2):\penalty0 e0262340, February 2023.
\newblock ISSN 1932-6203.
\newblock \doi{10.1371/journal.pone.0281380}.

\bibitem[Grochowicz et~al.(2024)Grochowicz, Van~Greevenbroek, and
  Bloomfield]{grochowicz-vangreevenbroek-ea-2024}
Aleksander Grochowicz, Koen Van~Greevenbroek, and Hannah~C Bloomfield.
\newblock Using power system modelling outputs to identify weather-induced
  extreme events in highly renewable systems.
\newblock \emph{Environmental Research Letters}, 19\penalty0 (5):\penalty0
  054038, May 2024.
\newblock ISSN 1748-9326.
\newblock \doi{10.1088/1748-9326/ad374a}.

\bibitem[Neuhausen et~al.(2023)Neuhausen, Rose, Bomke, Ferk, Wietschel,
  Pl{\"o}tz, Link, and Gnann]{neuhausen-rose-ea-2023}
J{\"o}rn Neuhausen, Philipp Rose, Jan-Hendrik Bomke, Felix Ferk, Martin
  Wietschel, Patrick Pl{\"o}tz, Steffen Link, and Till Gnann.
\newblock European {{Fleet Electrification}}.
\newblock Technical report, PwC, Fraufenhofer ISI, 2023.
\newblock URL
  \url{https://www.isi.fraunhofer.de/content/dam/isi/dokumente/cce/2023/2023-12-20_Strategy_Fraunhofer%20ISI%20-%20Fleet%20Electrification%20Study.pdf}.

\bibitem[Seibert et~al.(2024)Seibert, Kasten, Graichen, and
  Wissner]{seibert-kasten-ea-2024}
Dennis Seibert, Peter Kasten, Jakob Graichen, and Nora Wissner.
\newblock {{EU}} 2040 climate target: {{Contributions}} of the transport
  sector.
\newblock Technical report, Oeko-Institut, Berlin, 2024.
\newblock URL
  \url{https://www.oeko.de/fileadmin/oekodoc/EU2040ClimateTarget_potential-contributions-of-transport_final.pdf}.

\bibitem[{Transport \& Environment}(2023)]{transportenvironment-2023}
{Transport \& Environment}.
\newblock Modelling {{The Impact Of FuelEU Maritime On EU Shipping}}.
\newblock Technical report, 2023.
\newblock URL
  \url{https://www.transportenvironment.org/uploads/files/FuelEU-Maritime-Impact-Assessment.pdf}.

\bibitem[{The European Parliament and the Council of the European
  Union}(2023{\natexlab{a}})]{fueleu-2023}
{The European Parliament and the Council of the European Union}.
\newblock On the use of renewable and low-carbon fuels in maritime transport,
  2023{\natexlab{a}}.
\newblock URL \url{https://eur-lex.europa.eu/eli/reg/2023/1805}.

\bibitem[Caglayan et~al.(2020)Caglayan, Weber, Heinrichs, Lin{\ss}en, Robinius,
  Kukla, and Stolten]{caglayan-weber-ea-2020}
Dilara~Gulcin Caglayan, Nikolaus Weber, Heidi~U. Heinrichs, Jochen Lin{\ss}en,
  Martin Robinius, Peter~A. Kukla, and Detlef Stolten.
\newblock Technical potential of salt caverns for hydrogen storage in
  {{Europe}}.
\newblock \emph{International Journal of Hydrogen Energy}, 45\penalty0
  (11):\penalty0 6793--6805, February 2020.
\newblock ISSN 0360-3199.
\newblock \doi{10.1016/j.ijhydene.2019.12.161}.

\bibitem[{The European Parliament and the Council of the European
  Union}(2023{\natexlab{b}})]{eu-lulucf-2023}
{The European Parliament and the Council of the European Union}.
\newblock The inclusion of greenhouse gas emissions and removals from land use,
  land use change and forestry in the 2030 climate and energy framework, May
  2023{\natexlab{b}}.
\newblock URL \url{http://data.europa.eu/eli/reg/2018/841/2023-05-11/eng}.

\bibitem[{European Commission}(2024{\natexlab{c}})]{europeancommission-2024c}
{European Commission}.
\newblock {{EU Climate Action Progress Report}} 2024.
\newblock Technical Report COM/2024/498, 2024{\natexlab{c}}.
\newblock URL
  \url{https://eur-lex.europa.eu/legal-content/EN/TXT/?uri=CELEX%3A52024DC0498&qid=1730826894109}.

\bibitem[Pineda and Morales(2018)]{pineda-morales-2018}
S.~Pineda and J.~M. Morales.
\newblock Chronological {{Time-Period Clustering}} for {{Optimal Capacity
  Expansion Planning With Storage}}.
\newblock \emph{IEEE Transactions on Power Systems}, 33\penalty0 (6):\penalty0
  7162--7170, November 2018.
\newblock ISSN 1558-0679.
\newblock \doi{10.1109/TPWRS.2018.2842093}.

\bibitem[Kotzur et~al.(2018)Kotzur, Markewitz, Robinius, and
  Stolten]{kotzur-markewitz-ea-2018a}
Leander Kotzur, Peter Markewitz, Martin Robinius, and Detlef Stolten.
\newblock Impact of different time series aggregation methods on optimal energy
  system design.
\newblock \emph{Renewable Energy}, 117:\penalty0 474--487, March 2018.
\newblock ISSN 0960-1481.
\newblock \doi{10.1016/j.renene.2017.10.017}.

\bibitem[Pedersen et~al.(2021)Pedersen, Victoria, Rasmussen, and
  Andresen]{pedersen-victoria-ea-2021}
Tim~T. Pedersen, Marta Victoria, Morten~G. Rasmussen, and Gorm~B. Andresen.
\newblock Modeling all alternative solutions for highly renewable energy
  systems.
\newblock \emph{Energy}, 234:\penalty0 121294, November 2021.
\newblock ISSN 03605442.
\newblock \doi{10.1016/j.energy.2021.121294}.

\bibitem[Finke et~al.(2024)Finke, Weber, and Bertsch]{finke-weber-ea-2024}
Jonas Finke, Christoph Weber, and Valentin Bertsch.
\newblock Linking {{Modelling}} to {{Generate Alternatives}}, {{Multi-Objective
  Optimisation}} and {{Market Equilibria}} -- {{On}} the {{Economic
  Interpretation}} of {{Near-Cost-Optimal Solutions}} in {{Energy System
  Models}}, April 2024.

\end{thebibliography}
\renewcommand{\bibnumfmt}[1]{#1.}

\appendix
\clearpage
\onecolumn
\newgeometry{left=18mm, right=18mm}

\subsection*{Supplementary Information}

\setcounter{page}{1}

\begin{suptable}
  \centering
  \small
  \begin{tabular}{l>{\raggedleft\arraybackslash}p{1.9cm}p{3.5cm}p{7.5cm}}
    \toprule
    Ref. & \raggedright Green \ch{H2} prod. [Mt/a] & Model & Comment\\ \midrule
    Blanco et al.~\cite{blanco-nijs-ea-2018} & 0--120 & JRC-EU-TIMES & 95\% emissions reduction. No green hydrogen production in scenarios with unrestricted \ch{CO2} sequestration (amounting to 1400 Mt/a); 90--120 Mt/a in scenarios with no \ch{CO2} sequestration. \\ 
    Pickering et al.~\cite{pickering-lombardi-ea-2022} & 76--157 & Sector-coupled version of euro-calliope & Model without \ch{CO2} sequestration and fossil fuels; range is over 441 different near-optimal solutions at most 10\% more expensive than cost-optimal solution. \\
    Seck et al.~\cite{seck-hache-ea-2022} & 38--78 & Soft-linking of MIRET-EU \& Integrate Europe & Most of the green hydrogen production at ``offgrid'' sites (i.e. used at the site of production), reducing distribution costs. Tight upper limit of 1400 Mt/a on \ch{CO2} sequestration. \\
    Neumann et al.~\cite{neumann-zeyen-ea-2023} & 69--80 & PyPSA-Eur & Tight upper limit of 200 Mt/a on \ch{CO2} sequestration. \\
    Zeyen et al.~\cite{zeyen-victoria-ea-2023} & 114--119 & PyPSA-Eur & Tight upper limit of 200 Mt/a on \ch{CO2} sequestration; endogenously modelled technological learning for renewables and electrolysis leads to an increase in green hydrogen production. In additional analysis using a 2000 Mt/a limit on \ch{CO2} sequestration, green hydrogen production drops to $\sim$20 Mt/a. \\
    Béres et al.~\cite{beres-nijs-ea-2024} & 17--76 & Soft-linking of JRC-EU-TIMES and PLEXOS & Upper limit of 300--1000 Mt/a on \ch{CO2} sequestration depending on the scenario; assumes 50\% of total hydrogen demand must be met by imports. \\ 
    Schreyer et al.~\cite{schreyer-ueckerdt-ea-2024} & 8--76 & REMIND & Upper limit of 270 Mt/a on \ch{CO2} sequestration. Low green hydrogen production in this study is countered with zero-emissions energy imports of up to 1400 TWh/a. \\ 
    Kountouris et al.~\cite{kountouris-bramstoft-ea-2024} & 30--51 & Balmorel & \ch{CO2} sequestration not limited but peaks around 250 Mt/a in 2050; green hydrogen imports of up to about 200 TWh/a. Hydrogen demand largely determined exogenously; little competition between hydrogen and fossil fuels. \\
    Fleiter et al.~\cite{fleiter-fragoso-ea-2024} & 20--78 & Enertile & \ch{CO2} sequestration limited to cement production process emissions; the lower green hydrogen production scenarios rely on synthetic (green) fuel imports of up to \qty{1600}{TWh}. \\
    \bottomrule
  \end{tabular}
  \caption{
    Literature review of projections for European green hydrogen production by 2050, ordered by publication date. Results are compatible with a net-zero emissions target unless otherwise is indicated. Only studies including all major energy production and demand sectors (electricity, heat, transportation, industry) are included. Differences in projections are due to a variety of factors; most notably differing assumptions regarding \ch{CO2} sequestration potential, energy imports and, to a lesser extent, biomass availability. Other factors include cost assumptions and modelling methodology.
  }
  \label{suptab:green-h2-lit-review}
\end{suptable}

\begin{supfigure*}[h]
  \centering
  \includegraphics{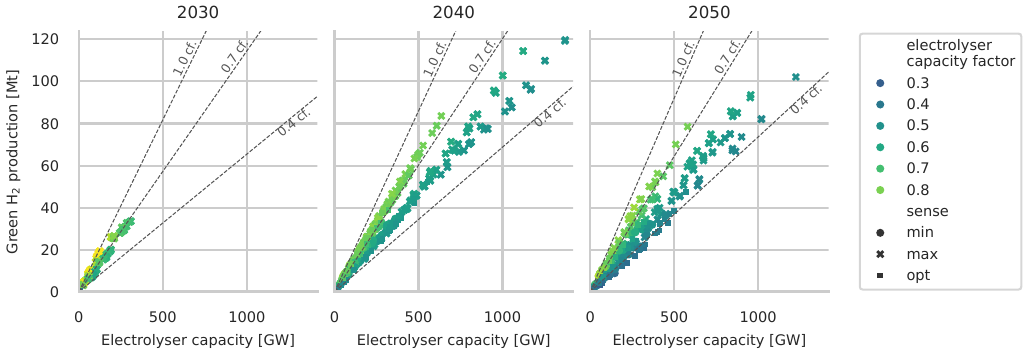}
  \caption{Relationship between total electrolyser capacity and total green hydrogen production at time horizons 2030, 2040 and 2050, including both cost-optimal and near-optimal results. Points are shaded by model-wide average annual electrolyser capacity factor, which in this plot is proportional to the ratio between the $x$- and $y$-coordinates of each point.}
  \label{supfig:h2-vs-elec-cap}
\end{supfigure*}

\begin{supfigure*}[h]
  \centering
  \includegraphics{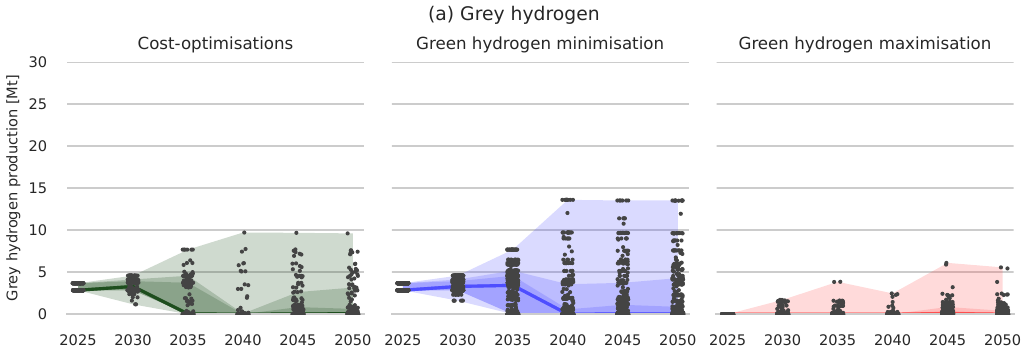}
  \includegraphics{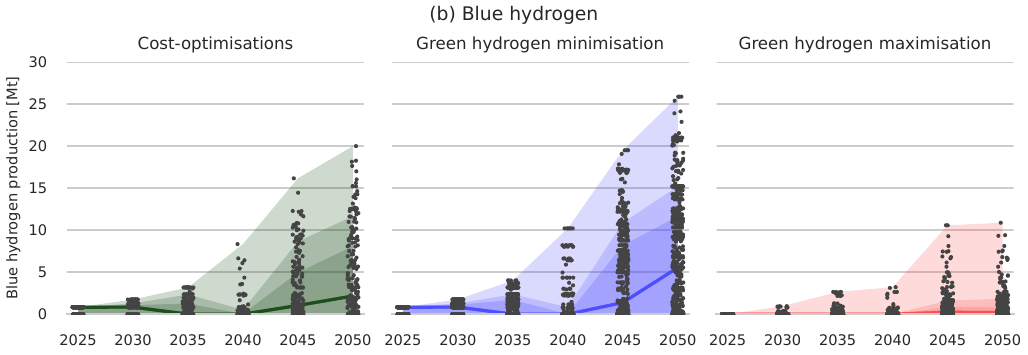}
  \includegraphics{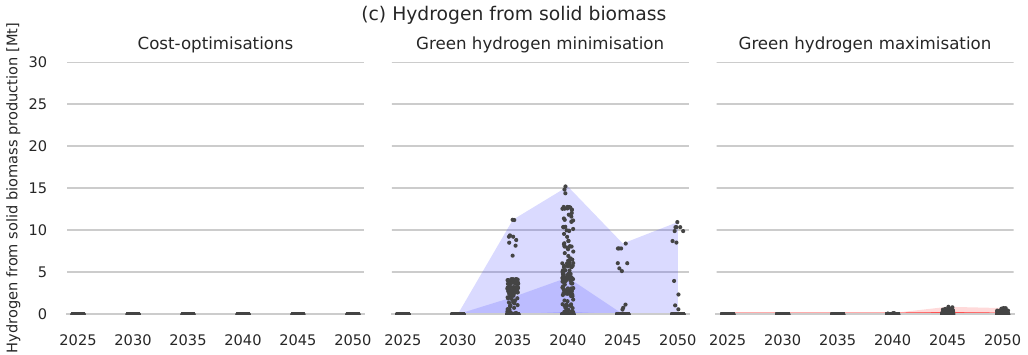}
  \caption{
    Evolution of total hydrogen produced by (a) steam methane reformation (b) steam methane reformation with carbon capture and (c) solid biomass.
    In each plot and time horizon, individual model results are plotted with black dots, while the solid line indicates the median of all model results.
    Meanwhile, 75th, 90th and 100th percentile ranges are shaded.
    Plotted model results include all scenarios and slack levels.
    These forms of hydrogen production are the only production pathways apart from green (i.e. electrolytic) hydrogen that are used in model.
    Ammonia cracking is also an option, but is never used.
  }
  \label{supfig:h2prod-by-tech}
\end{supfigure*}

\begin{supfigure*}[h]
  \centering
  \begin{minipage}{17.4cm}
    \begin{minipage}[b]{0.48\textwidth}
      \centering
      \includegraphics{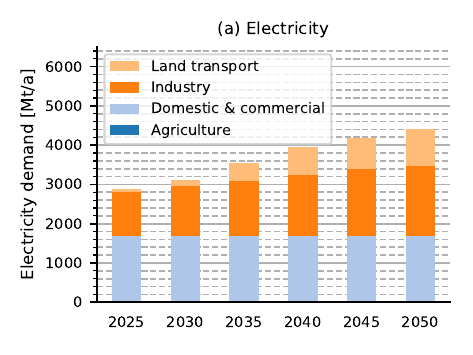}
    \end{minipage}
    \hfill
    \begin{minipage}[b]{0.48\textwidth}
      \centering
      \includegraphics{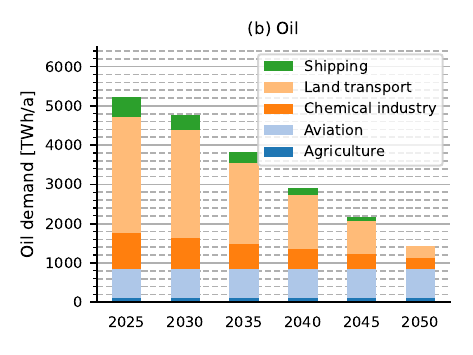}
    \end{minipage}
  \end{minipage}
  \caption{The two energy sectors with major exogenously fixed demand: electricity (a) and oil (b). Note that these figures \emph{do not} include demand for electricity and oil which is subject to endogenous optimisation such as heating, power-to-X, storage, etc.}
  \label{supfig:demand-fixed-major}
\end{supfigure*}

\begin{supfigure*}[h]
  \centering
  \includegraphics{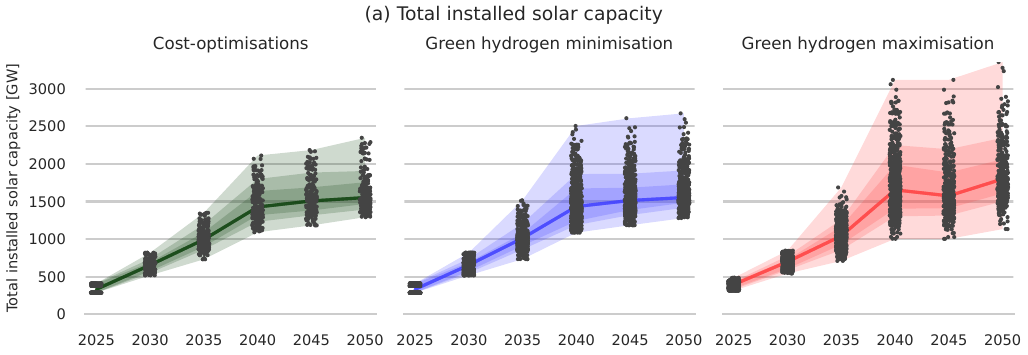}
  \includegraphics{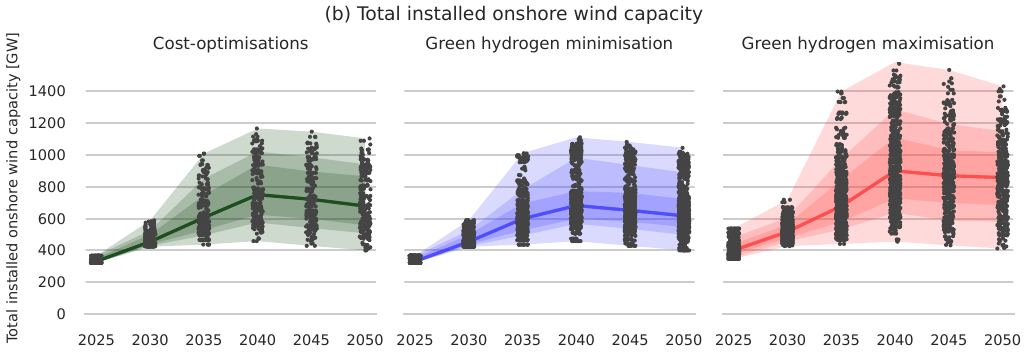}
  \includegraphics{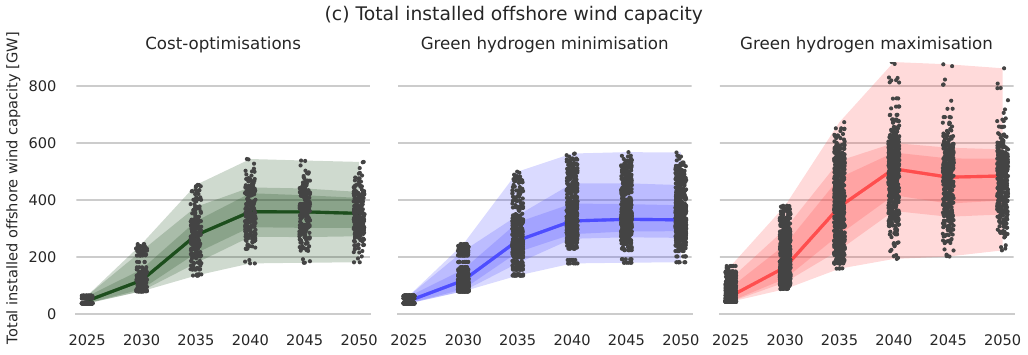}
  \caption{
    Evolution of total installed renewable capacity over time, in cost-optimisations as well as model runs where green hydrogen production is minimised and maximised.
    In each plot and time horizon, individual model results are plotted with black dots, while the solid line indicates the median of all model results.
    Meanwhile, 75th, 90th and 100th percentile ranges are shaded.
    Note that there are no scaling constraints on renewable capacities in the model.
  }
  \label{supfig:renewables}
\end{supfigure*}

\begin{supfigure*}[h]
  \centering
  \includegraphics{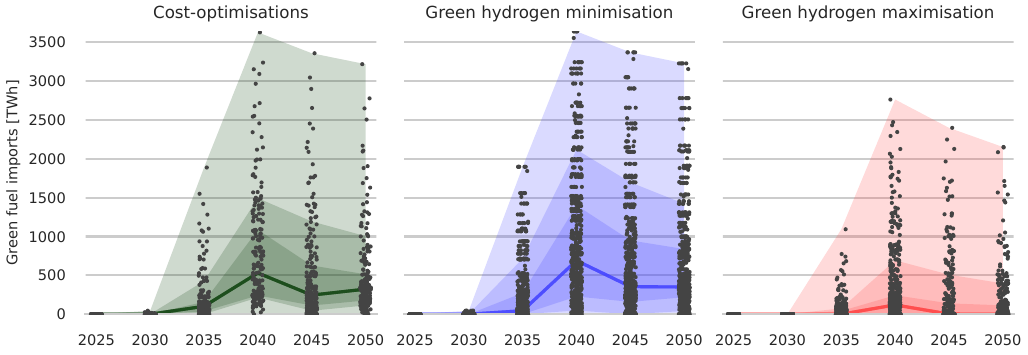}
  \caption{
    Evolution of total annual green fuel imports over time, in cost-optimisations as well as model runs where green hydrogen production is minimised and maximised.
    In each plot and time horizon, individual model results are plotted with black dots, while the solid line indicates the median of all model results.
    Meanwhile, 75th, 90th and 100th percentile ranges are shaded.}
  \label{supfig:imports}
\end{supfigure*}

\begin{supfigure*}[h]
  \centering
  \includegraphics{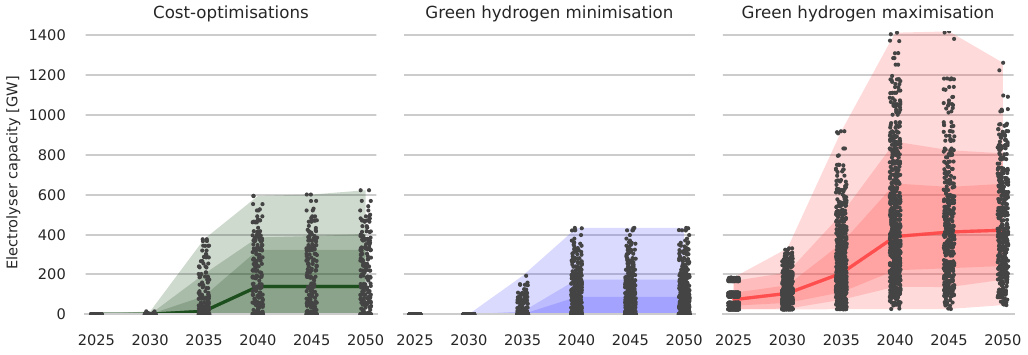}
  \caption{
    Evolution of total installed electrolyser capacity over time, in cost-optimisations as well as model runs where green hydrogen production is minimised and maximised.
    In each plot and time horizon, individual model results are plotted with black dots, while the solid line indicates the median of all model results.
    Meanwhile, 75th, 90th and 100th percentile ranges are shaded.
    Note that there are no scaling constraints on electrolyser capacity in the model.
  }
  \label{supfig:elec-cap}
\end{supfigure*}

\begin{supfigure*}[h]
  \centering
  \includegraphics{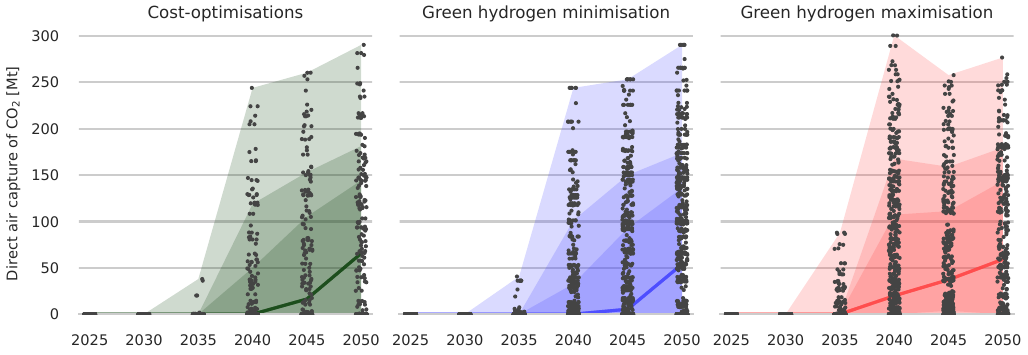}
  \caption{
    Evolution of total direct air capture (DAC) of \ch{CO2} over time, in cost-optimisations as well as model runs where green hydrogen production is minimised and maximised.
    In each plot and time horizon, individual model results are plotted with black dots, while the solid line indicates the median of all model results.
    Meanwhile, 75th, 90th and 100th percentile ranges are shaded.
    Note that there are no scaling constraints on DAC capacity in the model.
  }
  \label{supfig:DAC}
\end{supfigure*}

\begin{supfigure*}[h]
  \centering
  \includegraphics{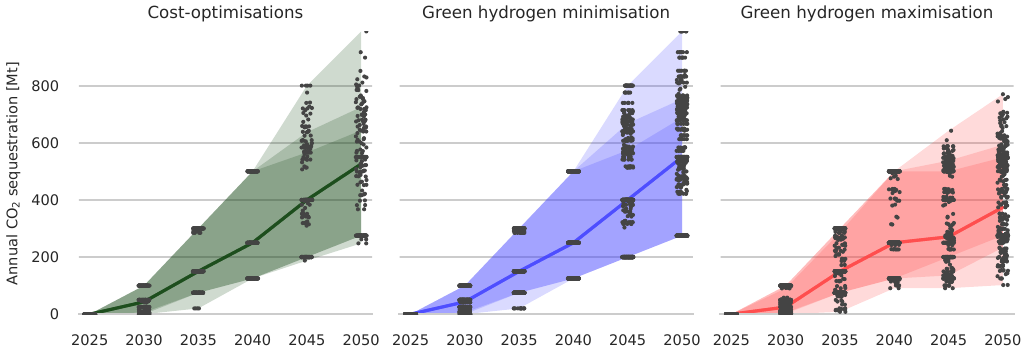}
  \caption{
    Evolution of total annual \ch{CO2}-sequestration over time, in cost-optimisations as well as model runs where green hydrogen production is minimised and maximised.
    In each plot and time horizon, individual model results are plotted with black dots, while the solid line indicates the median of all model results.
    Meanwhile, 75th, 90th and 100th percentile ranges are shaded. 
    For cost-optimal and green hydrogen-minimising results, the three different limits on \ch{CO2}-sequestration can clearly be seen at most time horizons as most model runs tend to fully use all available sequestration capacity.}
  \label{supfig:sequestration}
\end{supfigure*}

\begin{supfigure*}[h]
  \centering
  \includegraphics{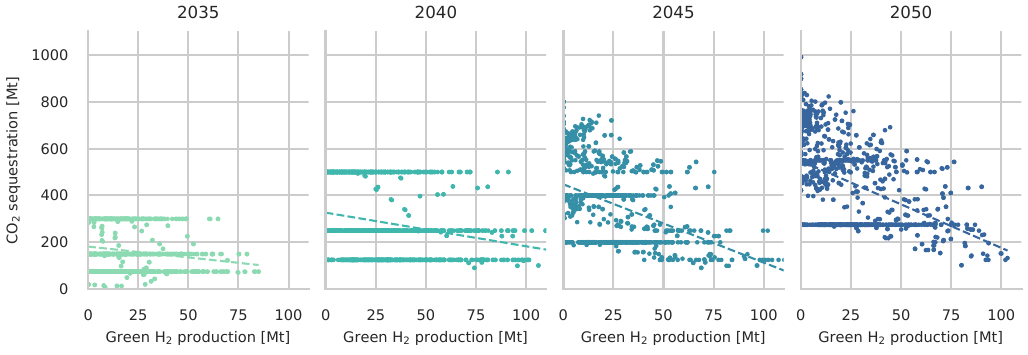}
  \caption{
    Scatter-plot of annual green hydrogen production against annual \ch{CO2} sequestration at time horizons 2035 -- 2050. Lines of best fit are also plotted.
    Clusters of data points at horizontal lines representing various limits to \ch{CO2} sequestration (levels (a), (b) and (c) of the CCS potential setting) are clearly visible.
  }
  \label{supfig:h2-vs-co2seq}
\end{supfigure*}

\begin{supfigure*}[h]
  \centering
  \includegraphics{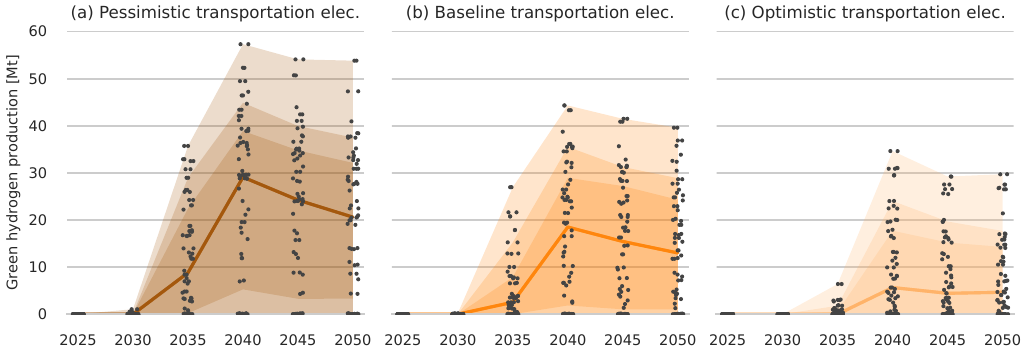}
  \caption{
    Evolution of cost-optimal total annual green hydrogen production over time, in pessimistic, baseline and optimistic land transportation electrification scenarios.
    In each plot and time horizon, individual model results are plotted with black dots, while the solid line indicates the median of all model results.
    Meanwhile, 75th, 90th and 100th percentile ranges are shaded.
  }
  \label{supfig:h2prod-by-transportation}
\end{supfigure*}

\begin{supfigure*}[h]
  \centering
  \begin{minipage}{17.4cm}
    \begin{minipage}[b]{0.48\textwidth}
      \centering
      \includegraphics{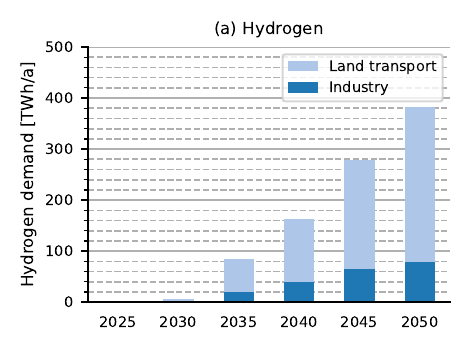}
    \end{minipage}
    \hfill
    \begin{minipage}[b]{0.48\textwidth}
      \centering
      \includegraphics{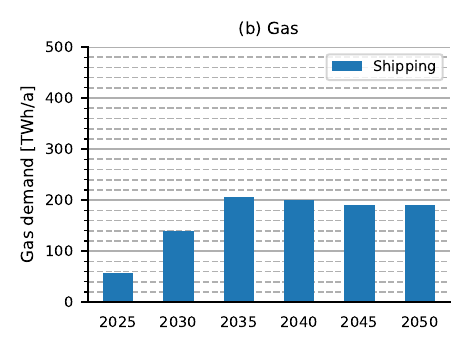}
    \end{minipage}
    \\[1ex]
    \begin{minipage}[b]{0.48\textwidth}
      \centering
      \includegraphics{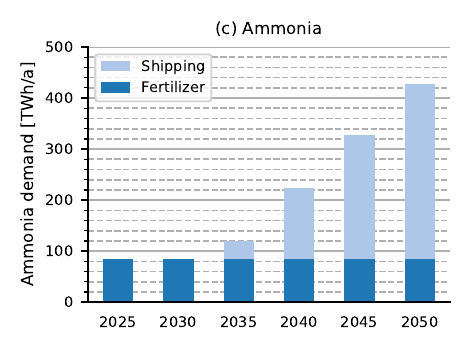}
    \end{minipage}
    \hfill
    \begin{minipage}[b]{0.48\textwidth}
      \centering
      \includegraphics{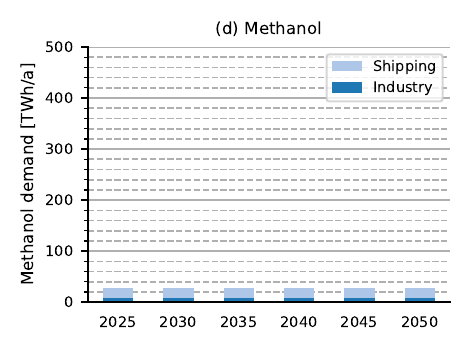}
    \end{minipage}
  \end{minipage}
  \caption{
    The remaining energy sectors with exogenously fixed demand: hydrogen (a), gas (b), ammonia (c) and methanol (d). Note that these figures \emph{do not} include demand which is subject to endogenous optimisation such as heating, power-to-X, storage, etc.; see \cref{supfig:demand-total-h2} for an overview of total demand for hydrogen, including endogenously optimised sectors.
  }
  \label{supfig:demand-fixed-minor}
\end{supfigure*}

\begin{supfigure*}[tbph]
  \centering
  \begin{minipage}{17cm}
    \begin{minipage}[b]{0.48\textwidth}
      \centering
      \includegraphics[width=\textwidth]{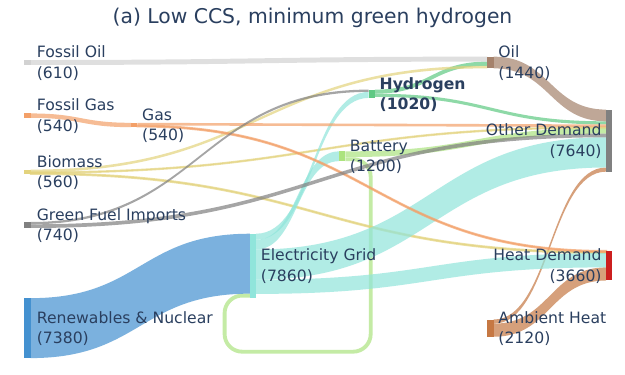}
    \end{minipage}
    \hfill
    \begin{minipage}[b]{0.48\textwidth}
      \centering
      \includegraphics[width=\textwidth]{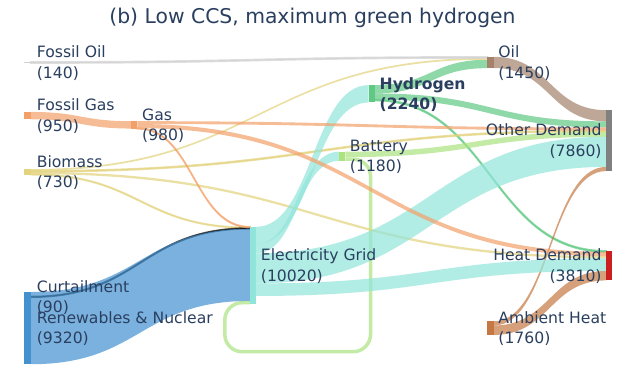}
    \end{minipage}
    \\[1ex]
    \begin{minipage}[b]{0.48\textwidth}
      \centering
      \includegraphics[width=\textwidth]{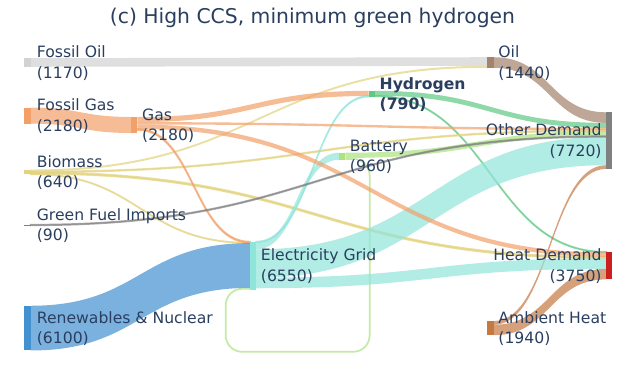}
    \end{minipage}
    \hfill
    \begin{minipage}[b]{0.48\textwidth}
      \centering
      \includegraphics[width=\textwidth]{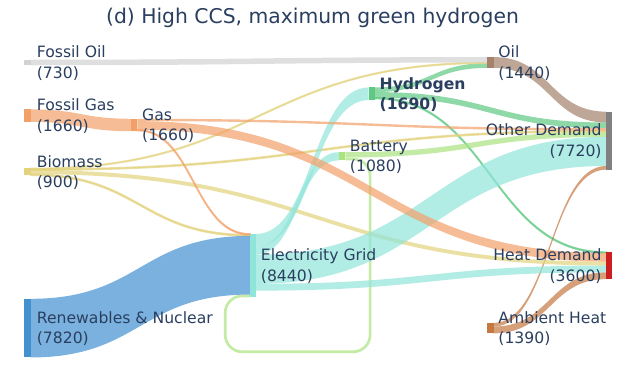}
    \end{minipage}
  \end{minipage}
  \caption{
    Energy flows in the same scenarios as in \cref{fig:sankey}, but shown in 2050 instead of 2040. All panels show max- and minimisation of green hydrogen production at a 5\% total system cost slack except panel (d), where the corresponding result for a 2\% slack is given since the optimisation for 5\% slack failed to converge.
  }
  \label{supfig:sankey-2050}
\end{supfigure*}

\begin{supfigure*}[tbph]
  \centering
  \begin{minipage}{17cm}
    \begin{minipage}[b]{0.48\textwidth}
      \centering
      \includegraphics[width=\textwidth]{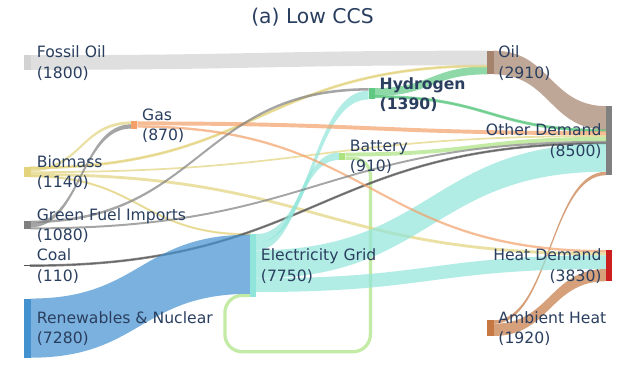}
    \end{minipage}
    \hfill
    \begin{minipage}[b]{0.48\textwidth}
      \centering
      \includegraphics[width=\textwidth]{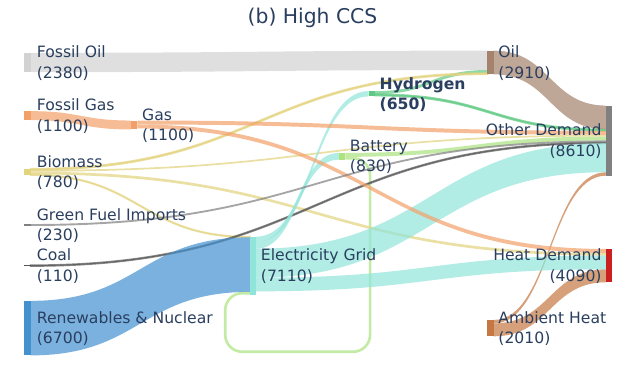}
    \end{minipage}
  \end{minipage}
  \caption{
    Energy flows in cost-optimal model results for the same scenarios as in \cref{fig:sankey}.
  }
  \label{supfig:sankey-opt-2040}
\end{supfigure*}

\begin{supfigure*}[tbph]
  \centering
  \begin{minipage}{17cm}
    \begin{minipage}[b]{0.48\textwidth}
      \centering
      \includegraphics[width=\textwidth]{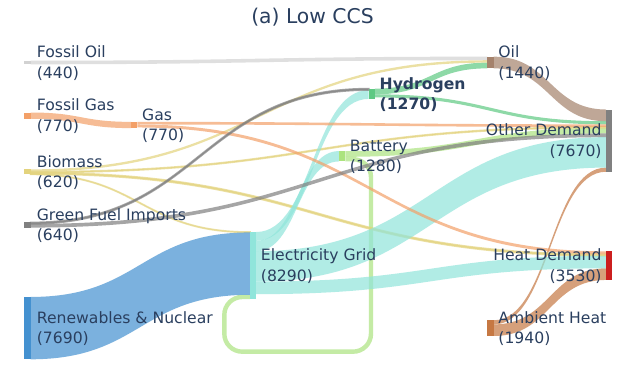}
    \end{minipage}
    \hfill
    \begin{minipage}[b]{0.48\textwidth}
      \centering
      \includegraphics[width=\textwidth]{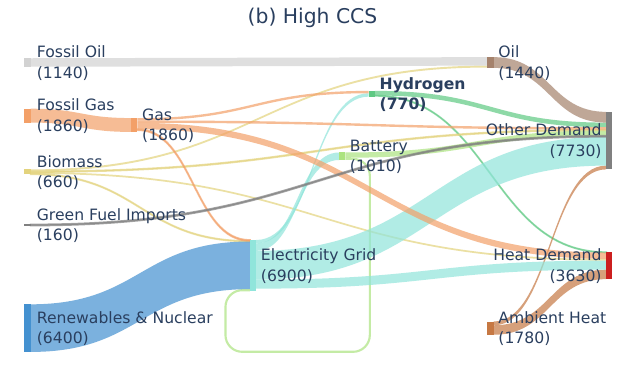}
    \end{minipage}
  \end{minipage}
  \caption{
    Energy flows in cost-optimal model results for the same scenarios as in \cref{fig:sankey}, but in 2050 instead of 2040.
  }
  \label{supfig:sankey-opt-2050}
\end{supfigure*}

\begin{supfigure*}[h]
  \centering
  \includegraphics{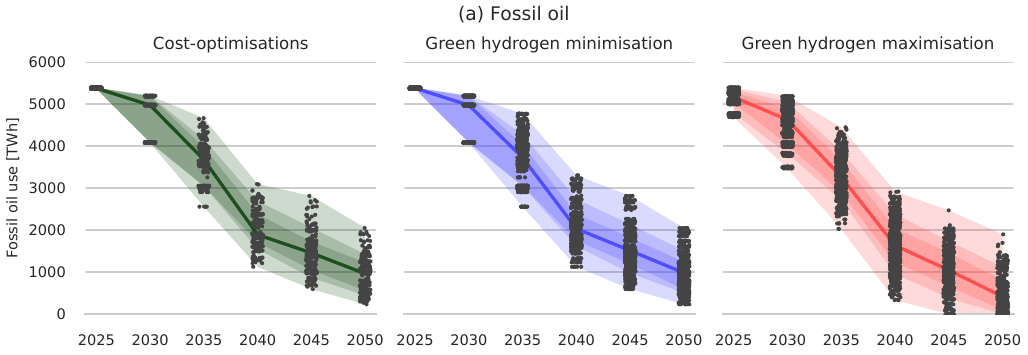}
  \includegraphics{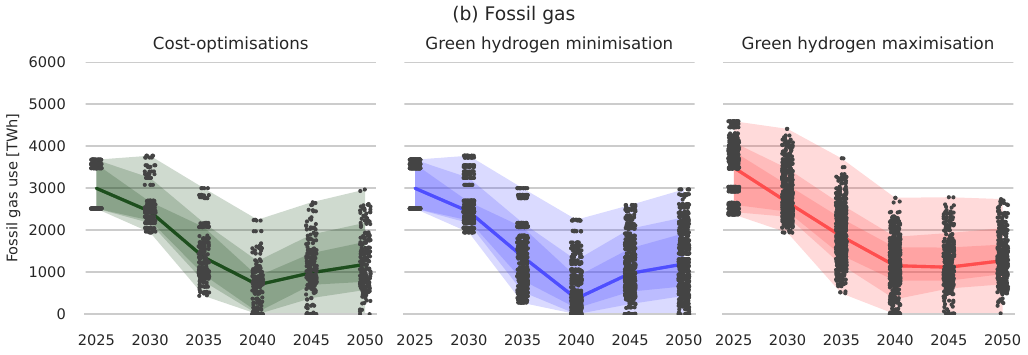}
  \caption{
    Evolution of total consumption of (a) fossil oil and (b) fossil gas by time horizon and cost-optimisations as well as green hydrogen min- and maximisations.
    In each plot and time horizon, individual model results are plotted with black dots, while the solid line indicates the median of all model results.
    Meanwhile, 75th, 90th and 100th percentile ranges are shaded.
  }
  \label{supfig:fossil-fuels}
\end{supfigure*}

\begin{supfigure*}[h]
  \centering
  \includegraphics{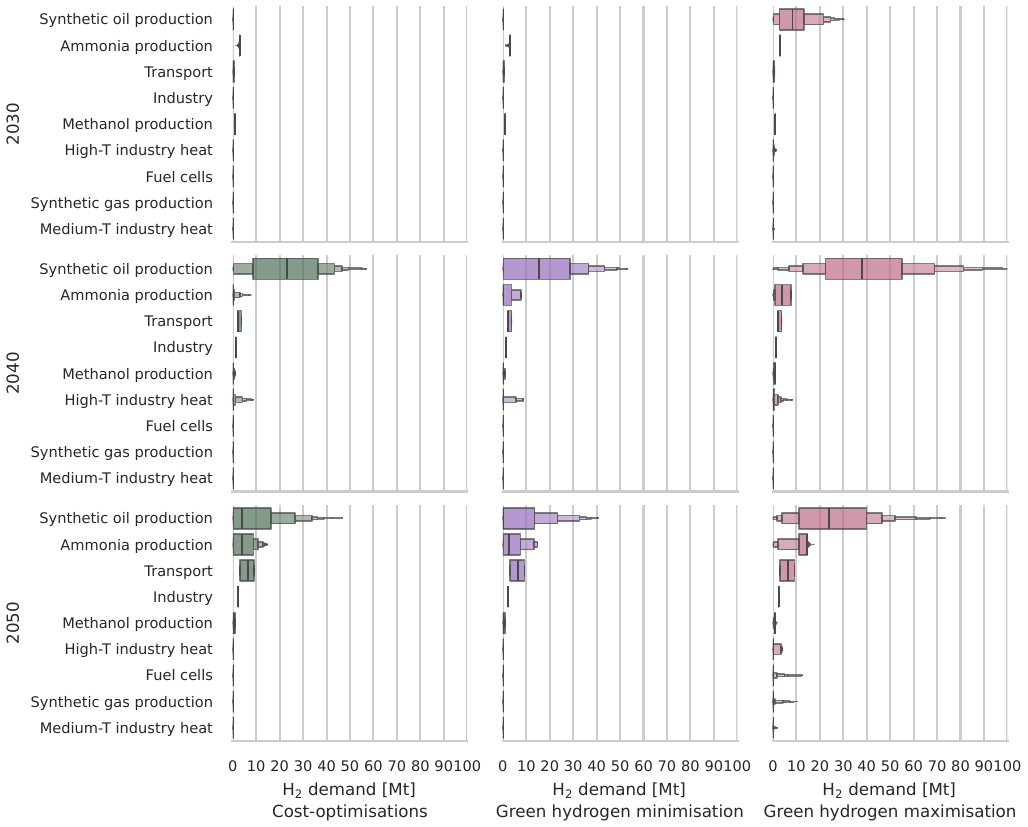}
  \caption{
    Distributions of types of hydrogen demand/use by planning horizon and optimisation sense.
    Of all the uses for hydrogen, the ``Industry'' and ``Transport'' (representing land transport) categories are exogenously fixed (see also \cref{supfig:demand-fixed-minor}), while all other use of hydrogen is endogenously optimised.
    The type of plot used is a letter-value plot \cite{letter-value-plot}; see also \cref{fig:pathways} for a description of this type of plot.
  }
  \label{supfig:demand-total-h2}
\end{supfigure*}

\begin{supfigure*}[h]
  \centering
  \includegraphics{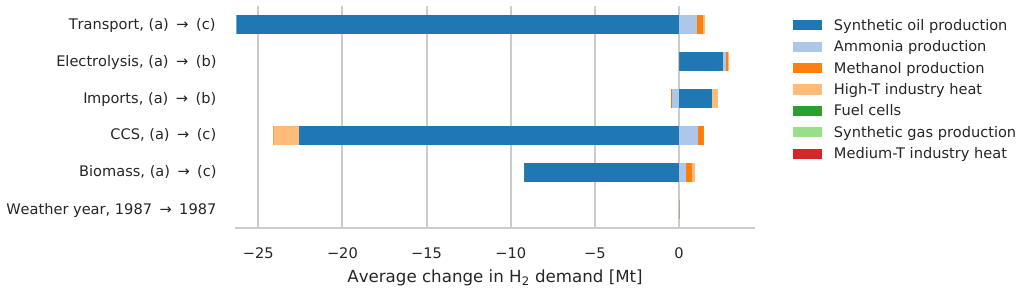}
  \caption{
    Change in hydrogen demand (of the demand that is not exogenously fixed) from one setting to another, on average across all scenarios.
    In other words, the bars show the difference in average hydrogen demand (by category) between all scenarios with level (a) and level (b)/(c) in in the six different scenario settings.
  }
  \label{supfig:demand-scenario-changes}
\end{supfigure*}

\begin{supfigure*}[h]
  \centering
  \includegraphics{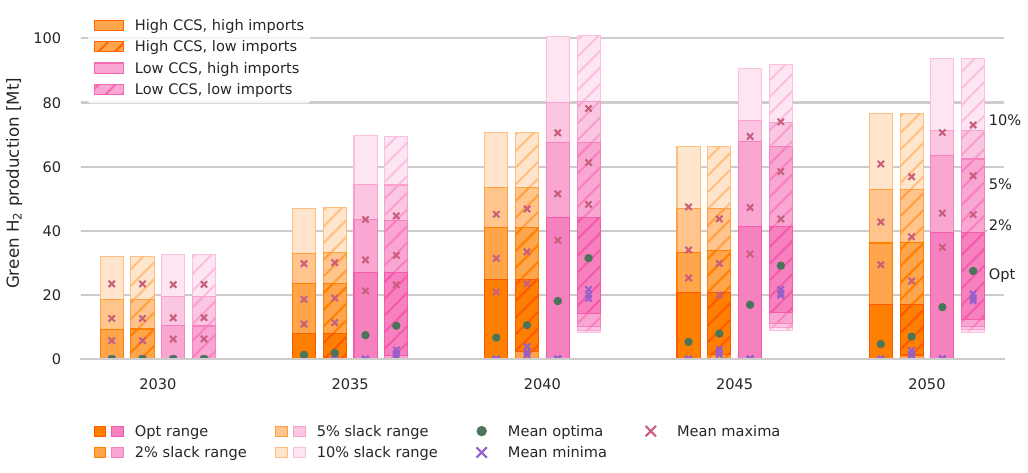}
  \caption{
    The ranges of pathways for European green hydrogen production under a combination of CCS and green fuel imports scenarios, assuming baseline transportation electrification.
    Here, the ``Low CCS'' and ``High CCS'' labels correspond to CCS potential levels (a) and (c), respectively; the ``low imports'' and ``high imports'' labels correspond to green fuel import levels (a) and (b), respectively (Methods).
    The different shades show the outer ranges across near-optimal results at the 10\%, 5\% and 2\% total system cost slack levels (from minimisation to maximisation of green hydrogen production), as well as the range of cost-optimal results.
    Markers show the mean of minimal, optimal and maximum green hydrogen production in across each scenario, at slack levels 2\%, 5\% and 10\% for minima and maxima.
    For example, among scenarios with ``Low CCS, low imports'' assumptions (and baseline transportation electrification), maximising green hydrogen production in 2050 with a 5\% total system cost slack leads to just under \qty{60}{Mt} on average (the annotated value on the right edge of the main plot marked with a cross), and about \qty{70}{Mt} at most across these scenarios (the height of the 5\%-shaded bar).
  }
  \label{supfig:CCS-imports}
\end{supfigure*}

\begin{supfigure*}[h]
  \centering
  \includegraphics[width=17.2cm]{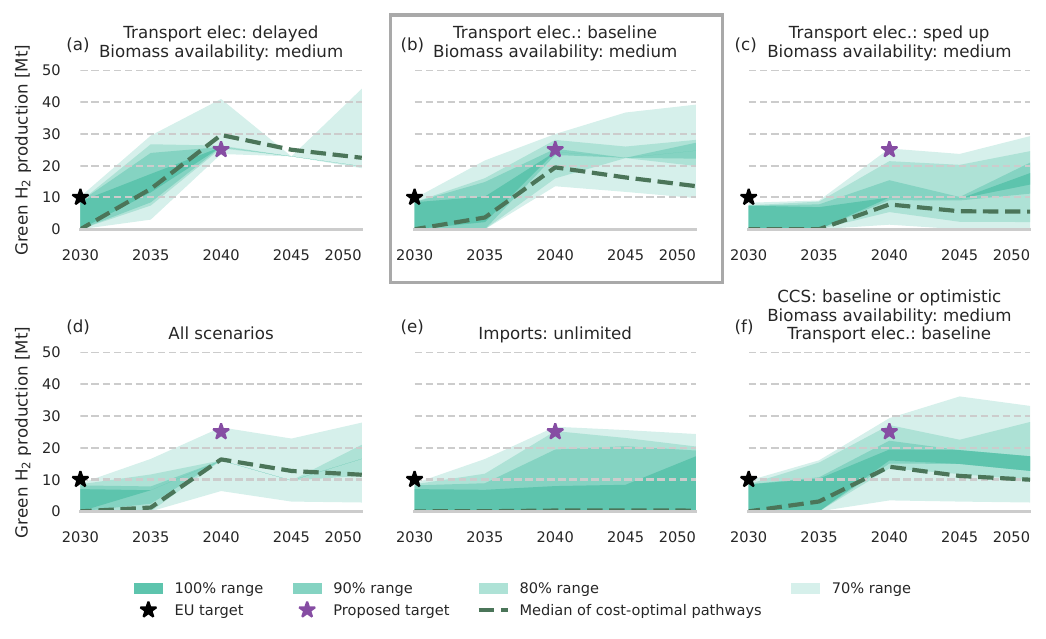}
  
  \vspace{1cm}
  
  \includegraphics[width=17.2cm]{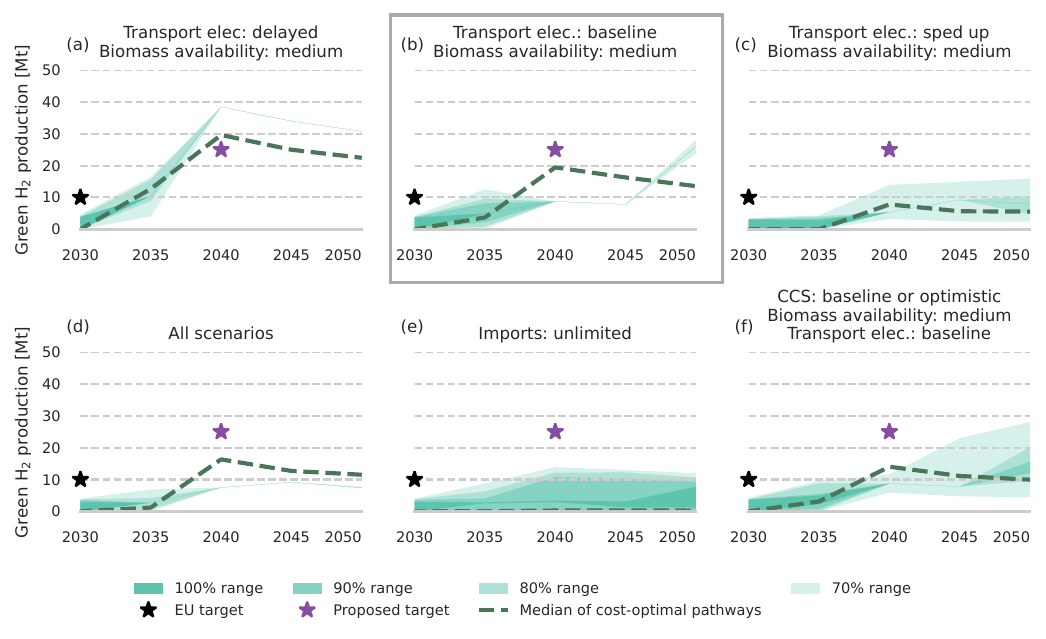}
  \caption{
    Robust corridors of green hydrogen production, analogous to \cref{fig:robustness}, for total system cost slack levels of $5\%$ (top) and $2\%$ (bottom).
  }
  \label{supfig:robustness-slack-levels}
\end{supfigure*}

\begin{supfigure*}[h]
  \centering
  \includegraphics{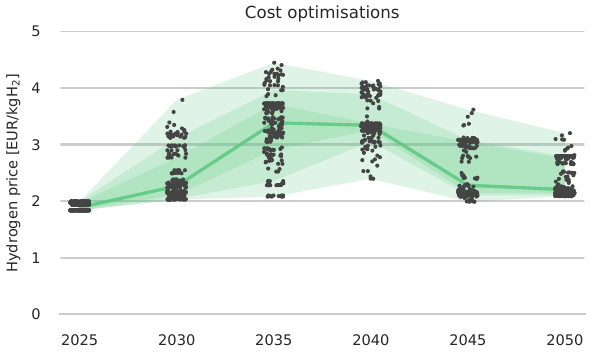}
  \caption{
    Hydrogen price in EUR/kgH$_2$ over time. This price is the overall hydrogen price (which may include production of hydrogen with natural gas and biomass with and without carbon capture), obtained as the average over time and space of the shadow prices of hydrogen buses in the optimisation model. In each plot and time horizon, individual model results are plotted with black dots, while the solid line indicates the median of all model results.
    Meanwhile, 75th, 90th and 100th percentile ranges are shaded.
  }
  \label{supfig:h2price}
\end{supfigure*}

\begin{supfigure*}[h]
  \centering
  \includegraphics[width=17.2cm]{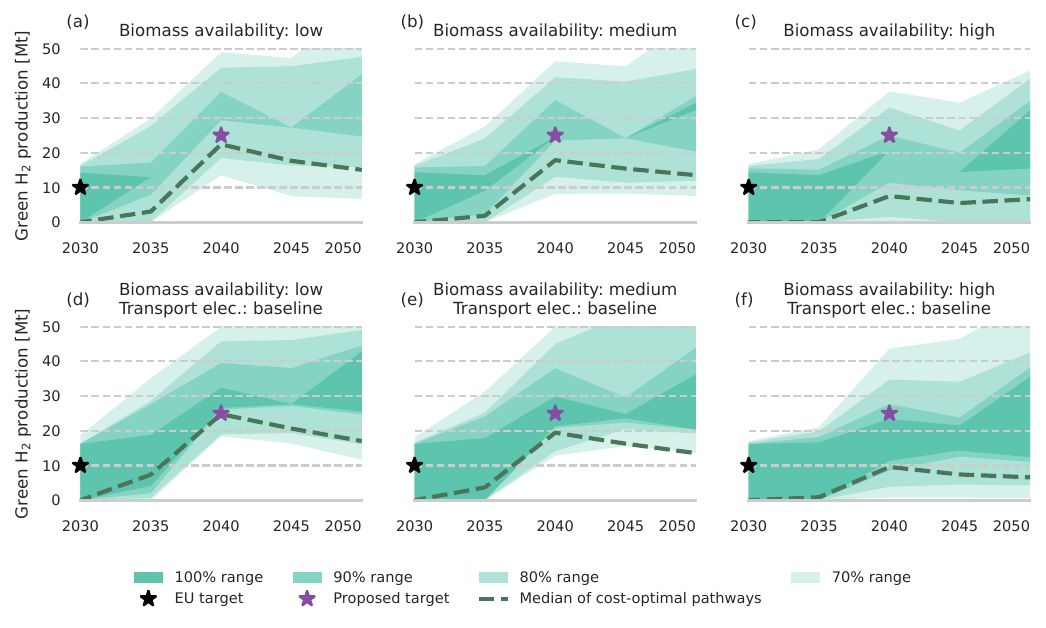}
  \caption{
    Robust corridors of green hydrogen production, analogous to \cref{fig:robustness}, with sets of scenarios including on low, medium and high biomass availability.
    In the bottom row (panels (d)--(f)), we restrict the scenario selections further to those with the baseline transportation electrification setting; as such, panel (e) is identical to panel (b) in \cref{fig:robustness}.
  }
  \label{supfig:robustness-inc-biomass}
\end{supfigure*}

\begin{supfigure*}[h]
  \centering  \includegraphics[width=17.2cm]{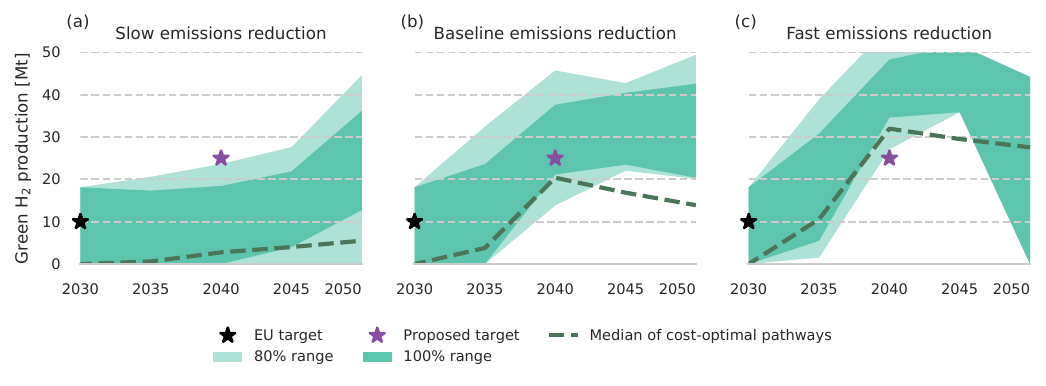}
  \caption{
    Robust corridors of European green hydrogen production under different emission reduction schemes. In the ``slow'', ``baseline'' and ``fast'' emission reduction scenarios, emissions have to be reduced by 80\%, 90\% and 95\% by 2040, respectively. Emission limits for 2030 (55\% reduction) and 2050 (100\% reduction, i.e. net zero) are held constant; limits for 2035 and 2045 are linearly interpolated.
  }
  \label{supfig:robustness-emission-reductions}
\end{supfigure*}

\begin{supfigure*}[h]
  \centering
  \includegraphics{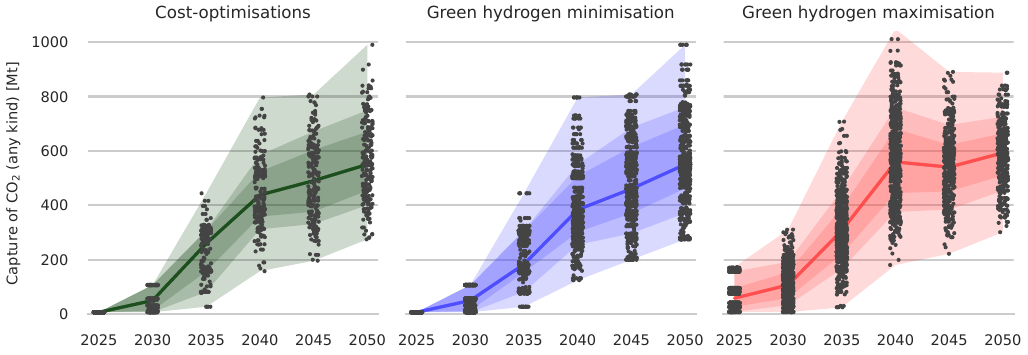}
  \caption{
    Evolution of total carbon capture of any kind over time, in cost-optimisations as well as model runs where green hydrogen production is minimised and maximised.
    This includes carbon capture at point emission sources in industry, power plants (including combined heat and power plants run on biomass) as well as direct air capture.
    In each plot and time horizon, individual model results are plotted with black dots, while the solid line indicates the median of all model results.
    Meanwhile, 75th, 90th and 100th percentile ranges are shaded.
  }
  \label{supfig:captured-co2}
\end{supfigure*}

\begin{supfigure*}[h]
  \centering
  \includegraphics{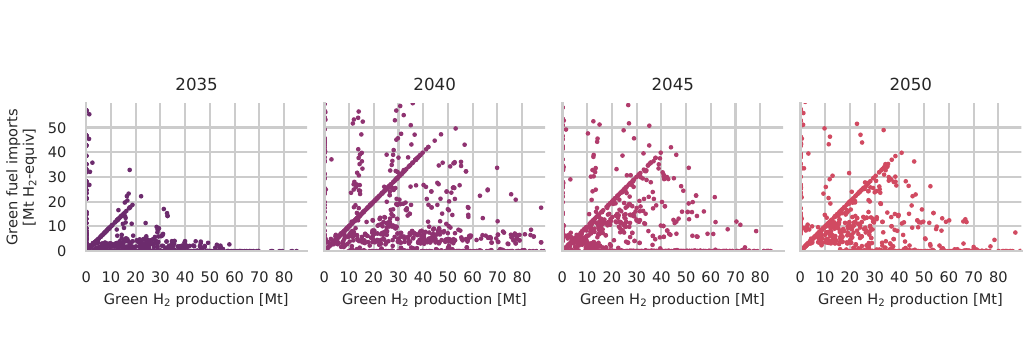}
  \caption{Scatter-plot of annual green hydrogen production against annual green fuel imports at time horizons 2035 -- 2050, including cost-optimisations as well as near-optimal solutions. Green fuel imports include green hydrogen, ammonia, methanol, synthetic gas and synthetic oil, but are here plotted in Mt \ch{H2}-equivalent units in order to facilitate direct comparison with green hydrogen production. The accumulation of points around the $y = x$ line indicate results in the restrictive imports scenario, where ``green fuel imports $\leq$ green hydrogen production'' constraint is binding.}
  \label{supfig:h2-vs-imports}
\end{supfigure*}

\begin{supfigure*}[h]
  \centering
  \includegraphics{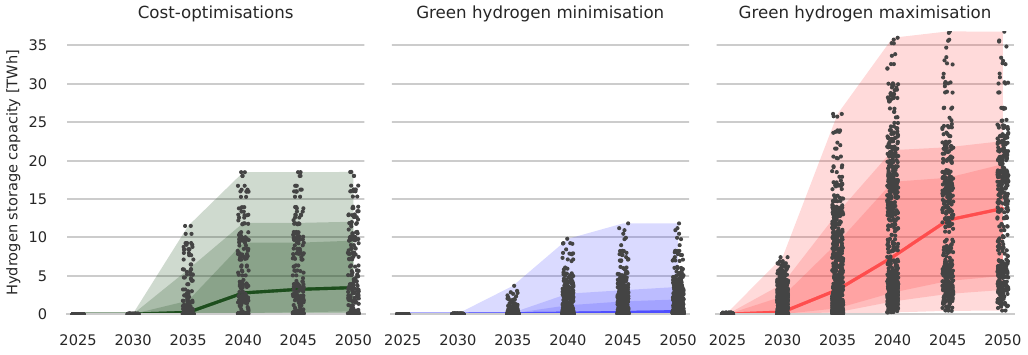}
  \caption{
    Evolution of total installed hydrogen storage capacity over time, in cost-optimisations as well as model runs where green hydrogen production is minimised and maximised.
    Hydrogen storage includes both underground salt cavern storage and overground compressed tank storage, but only salt cavern storage is ever used.
    In each plot and time horizon, individual model results are plotted with black dots, while the solid line indicates the median of all model results.
    Meanwhile, 75th, 90th and 100th percentile ranges are shaded.
}
  \label{supfig:h2-storage}
\end{supfigure*}

\begin{supfigure*}[h]
  \centering
  \includegraphics{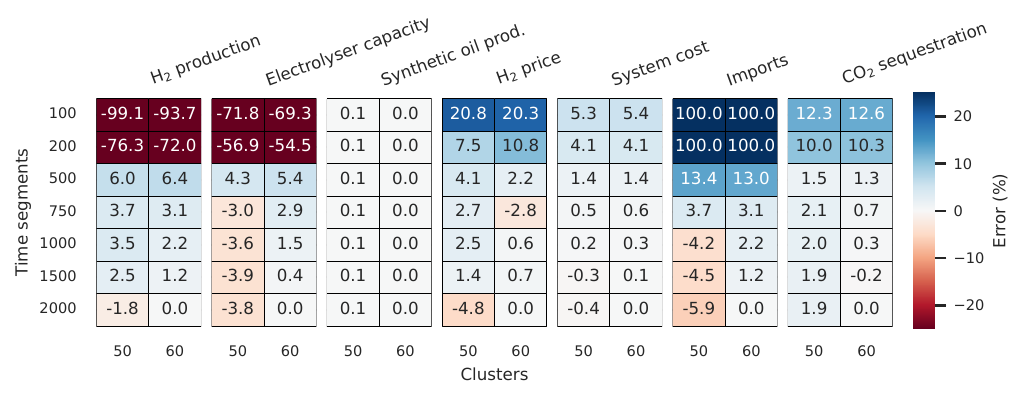}
  \caption{
    Relative percentage errors in selected model metrics between modelling runs of different resolutions.
    Shown here are the maximum relative errors across two different scenarios; setting (a) and (c) for CCS potential, with the other dimensions being medium biomass availability, limited green fuel imports, baseline electrolyser costs and baseline transportation electrification.
    All errors are relative to the 2000 time step, 60 cluster model (bottom right).
    Furthermore, the maximum is taken over cost-optimisations and green hydrogen min- and maximisations (with a 5\% total cost slack) as well as three planning horizons (2040, 2045 and 2050).
    (At earlier planning horizons, some metrics are nearly 0, leading to low absolute errors but very high relative errors.)
    As such, $2 \cdot 3 \cdot 3 = 18$ different comparisons can be made for each resolution.
    Out of these 18, the 2000 time step, 60 cluster model only solved successfully in 8 cases; the maximum relative errors across these 8 cases are shown.
  }
  \label{supfig:resolution-errors}
\end{supfigure*}

\end{document}